\documentclass[journal]{IEEEtran}

\usepackage{ifpdf}
\ifpdf
  \usepackage[pdftex]{graphicx}
  \usepackage[pdftex]{color}
  \usepackage[pdftex, hidelinks]{hyperref}
  \usepackage{epstopdf}
\else
  \usepackage[dvipdfmx]{graphicx}
  \usepackage[dvipdfmx]{xcolor}
  \usepackage[dvipdfmx, hidelinks]{hyperref}
\fi

\usepackage[caption=false,font=footnotesize]{subfig}
\usepackage{cite,url,color}
\usepackage{array,amsmath,amssymb,amsfonts,bm,mathtools}
\usepackage{algorithmic,algorithm}
\usepackage{booktabs,multirow}
\usepackage{threeparttable}
\usepackage{makecell}
\usepackage{diagbox}
\usepackage{fixltx2e}
\def\Argmax{\mathop{\rm argmax}}

\def\X{\mathbf{X}}
\def\Y{\mathbf{Y}}
\def\V{\mathbf{V}}
\def\H{\mathbf{H}}
\def\W{\mathbf{W}}
\def\Z{\mathbf{Z}}
\def\G{\mathbf{G}}
\def\P{\mathbf{P}}
\def\Q{\mathbf{Q}}

\def\p{\mathbf{p}}
\def\q{\mathbf{q}}
\def\w{\mathbf{w}}
\def\s{\mathbf{s}}
\def\x{\mathbf{x}}

\def\u{\mathbf{u}}

\def\g{\mathbf{g}}
\def\I{\mathbf{I}}

\def\mPsi{\bm\Psi}
\def\mPhi{\bm\Phi}

\def\zeros{\mathbf{0}}

\def\tr{\mathrm{T}}
\def\hr{\mathrm{H}}

\def\scr#1{\mbox{\scriptsize #1}}

\def\ComplexGaussian{\mathcal{N}_{\mathbb{C}}}

\def\wf{{\mbox{\tiny WF}}}
\def\wfone{{\mbox{\tiny WF1}}}
\def\mv{{\mbox{\tiny MV}}}

\graphicspath{{./figure/}}

\begin{document}

\title{Unsupervised Speech Enhancement Based on\\
Multichannel NMF-Informed Beamforming\\
for Noise-Robust Automatic Speech Recognition}

\fussy

\author{
Kazuki~Shimada,
Yoshiaki~Bando,~\IEEEmembership{Member,~IEEE,}
Masato~Mimura,
Katsutoshi~Itoyama,~\IEEEmembership{Member,~IEEE,}
Kazuyoshi~Yoshii,~\IEEEmembership{Member,~IEEE,}
and~Tatsuya~Kawahara,~\IEEEmembership{Fellow,~IEEE}
\thanks{
Manuscript received XXXX XX, 2018; 
revised XXXX XX, 2018; 
revised XXXX XX, 2018; 
accepted XXXX XX, 2019.
Date of publication XXXX XX, 2019; 
date of current version XXXX XX, 2019. 
This work was supported in part 
by JST ERATO No.~JPMJER1401. 
The associate editor coordinating the review of this manuscript 
and approving it for publication was XXXX XXXX
(Corresponding author: Kazuyoshi Yoshii).

K. Shimada, Y. Bando, M. Mimura, K. Itoyama, and T. Kawahara are 
with the Graduate School of Informatics, Kyoto University, Kyoto 606-8501, Japan
(email: shimada@sap.ist.i.kyoto-u.ac.jp, bando@sap.ist.i.kyoto-u.ac.jp, 
mimura@sap.ist.i.kyoto-u.ac.jp, itoyama@i.kyoto-u.ac.jp, kawahara@i.kyoto-u.ac.jp).

K.Yoshii is 
with the Graduate School of Informatics, Kyoto University, Kyoto 606-8501, Japan, 
and also with the RIKEN Center for Advanced Intelligence Project, Tokyo 103-0027, Japan 
(e-mail: yoshii@i.kyoto-u.ac.jp).

The final published version of this paper is available online
at http://ieeexplore.ieee.org.
}
}

\markboth{Accepted to IEEE/ACM TRANSACTIONS ON AUDIO, SPEECH, AND LANGUAGE PROCESSING}%
{K. Shimada \MakeLowercase{\textit{et al.}}: Unsupervised Speech Enhancement}

\IEEEpubid{
\begin{minipage}{\textwidth}\ \\[12pt]
\\
\end{minipage}
}


\maketitle

\begin{abstract}
This paper describes multichannel speech enhancement 
 for improving automatic speech recognition (ASR) in noisy environments.
Recently, 
 the minimum variance distortionless response (MVDR) beamforming has widely been used
 because it works well if the steering vector of speech 
 and the spatial covariance matrix (SCM) of noise are given.
To estimating such spatial information,
 conventional studies take a supervised approach 
 that classifies each time-frequency (TF) bin into noise or speech
 by training a deep neural network (DNN).
The performance of ASR, however, 
 is degraded in an unknown noisy environment.
To solve this problem,
 we take an unsupervised approach
 that decomposes each TF bin into the sum of speech and noise
 by using multichannel nonnegative matrix factorization (MNMF).
This enables us to accurately estimate the SCMs of speech and noise
 not from observed noisy mixtures but from separated speech and noise components.
In this paper
 we propose online MVDR beamforming 
 by effectively 
 initializing and incrementally updating the parameters of MNMF.
Another main contribution is
 to comprehensively investigate the performances of ASR
 obtained by various types of spatial filters, i.e., 
 time-invariant and variant versions of MVDR beamformers
 and those of rank-1 and full-rank multichannel Wiener filters,
 in combination with MNMF.
The experimental results showed that the proposed method 
 outperformed the state-of-the-art DNN-based beamforming method
 in unknown environments that did not match training data.
\end{abstract}


%
\IEEEpeerreviewmaketitle

\section{Introduction}
\label{sec:introduction}


\IEEEPARstart{M}{ultichannel} speech enhancement using a microphone array
 plays a vital role for distant automatic speech recognition (ASR) 
 in noisy environments.
A standard approach to multichannel speech enhancement
 is to use beamforming
 \cite{higuchi2017online,nakatani2017integrating,heymann2016nngev,
 erdogan2016mask,pertila2017microphone,wang2018multi,
 xiao2017time,ochiai2017multichannel,sainath2017multichannel,mimura2017combined}.
Given the spatial information of speech and noise, 
 we can emphasize the speech coming from one direction 
 and suppress the noise from the other directions 
 \cite{souden2010optimal,wang2018rank1arxiv,van1988beamforming,gannot2004speech,warsitz2007blind}.
This approach was empirically shown to achieve
 the significant improvement of ASR performance
 in the CHiME Challenge \cite{barker2015chime,yoshioka2015ntt,hori2015merl}.
There are many variants of beamforming
 such as 
 multichannel Wiener filtering (MWF) \cite{souden2010optimal,wang2018rank1arxiv}, 
 minimum variance distortionless response (MVDR) beamforming \cite{van1988beamforming}, 
 generalized sidelobe cancelling (GSC) \cite{gannot2004speech}, 
 and generalized eigenvalue (GEV) beamforming \cite{warsitz2007blind},
 which are all performed in the time-frequency (TF) domain.

\IEEEpubidadjcol

To calculate demixing filters for beamforming,
 the steering vector of speech 
 and the spatial covariance matrix (SCM) of noise should be estimated.
The steered response power phase transform (SRP-PHAT) \cite{loesch2010adaptive} 
 and the weighted delay-and-sum (DS) beamforming \cite{anguera2007beamformit}
 are not sufficiently robust to real environments \cite{barker2015chime}.
Recently, estimation of TF masks 
 has actively been investigated
 \cite{higuchi2017online,nakatani2017integrating,
 heymann2016nngev,erdogan2016mask,pertila2017microphone,wang2018multi,
 xiao2017time,ochiai2017multichannel},
 assuming that each TF bin of an observed noisy speech spectrogram 
 is classified into speech or noise.
The SCMs of speech and noise 
 are then calculated from the classified TF bins.
The steering vector of the target speech is obtained 
 as the principal component of the SCM of the speech 
 \cite{higuchi2017online,heymann2016nngev,nakatani2017integrating}.
For such binary classification,
 an unsupervised method 
 based on complex Gaussian mixture models (CGMMs) \cite{higuchi2017online}
 and a supervised method
 based on deep neural networks (DNNs)
 \cite{heymann2016nngev,erdogan2016mask,nakatani2017integrating,xiao2017time,
 ochiai2017multichannel,pertila2017microphone,wang2018multi}
 have been proposed.

Although DNN-based beamforming works well 
 in controlled experimental environments,
 it has two major problems in real environments.
One problem is that 
 the performance of ASR in unknown environments
 is often be considerably degraded 
 due to the overfitting to training data
 consisting of many pairs of noisy speech spectrograms 
 and ideal binary masks (IBMs) of speech. 
Although multi-condition training 
 with various kinds of noisy environments 
 mitigates the problem \cite{vincent2017analysis},
 it is still an open question 
 whether DNN-based beamforming works when a microphone array
 with different geometry and frequency characteristics 
 is used in unseen noisy environments.
The other problem
 is that the spatial features such as 
 inter-channel level and phase differences (ILDs and IPDs),
 which play an essential role 
 in conventional multichannel audio signal processing,
 are simply input to DNNs
 without considering the physical meanings 
 and generative processes of those features.

\begin{figure}[t]
\centering
\includegraphics[width=.98\linewidth]{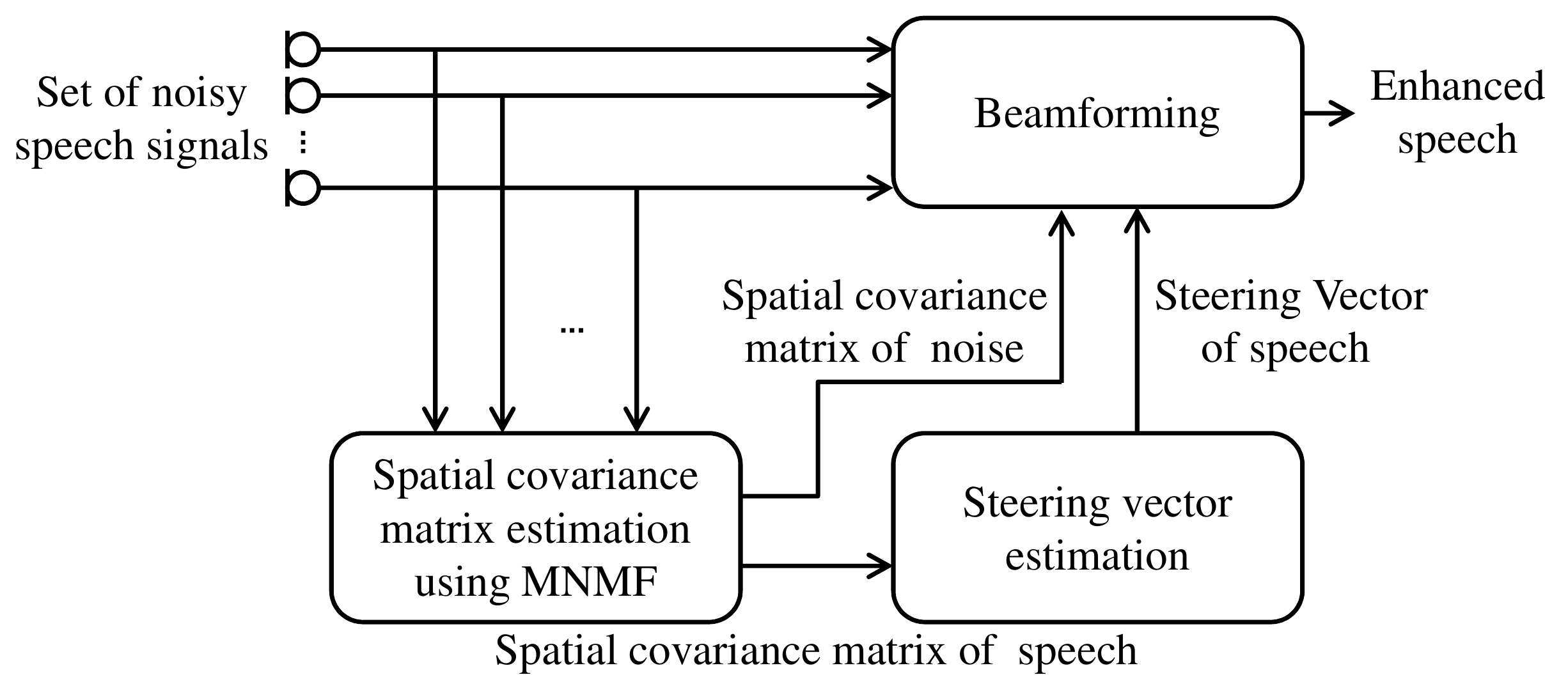}
\caption{The proposed approach to unsupervised speech enhancement 
based on a variant of beamforming
that calculates the SCMs of speech and noise
from the corresponding spectrograms obtained by MNMF.}
\label{fig:pm_framework_intro}
\end{figure}

To solve these problems,
 we recently proposed an unsupervised method 
 of speech enhancement \cite{shimada2018icassp}
 based on several types of beamforming 
 using the SCMs of speech and noise
 estimated by a blind source separation (BSS) method 
 called multichannel nonnegative matrix factorization (MNMF)\cite{sawada2013mnmf}
 (Fig.~\ref{fig:pm_framework_intro}).
Given the multichannel complex spectrograms of mixture signals, 
 MNMF can estimate the SCMs of multiple sources 
 ({\it i.e.}, speech and multiple noise sources)
 as well as represent the nonnegative power spectrogram of each source 
 as the product of two nonnegative matrices corresponding 
 to a set of basis spectra and a set of temporal activations.
The SCMs of speech and noise estimated by MNMF
 that {\it decomposes} each TF bin into the sum of speech {\it and} noise
 are expected to be more accurate 
 than those estimated by a CGMM\cite{higuchi2017online} 
 or a DNN\cite{heymann2016nngev,erdogan2016mask,pertila2017microphone,wang2018multi,
 nakatani2017integrating,xiao2017time,ochiai2017multichannel}
 that {\it classifies} each TF bin into speech {\it or} noise.
The unsupervised speech enhancement method is also expected to work well
 even in unknown environments for which there are no matched training data.
In this paper,
 we newly propose an online extension of MNMF-informed beamforming
 that can process the observed mixture signals in a streaming manner.

The main contribution of this paper
 is to describe the complete formulation 
 of the proposed method of MNMF-informed beamforming 
 and report comprehensive comparative experiments.
More specifically, 
 we test the combination of MNMF
 with various types of beamforming,
 {\it i.e.}, the time-variant and time-invariant versions 
 of full-rank MWF \cite{souden2010optimal}, 
 rank-1 MWF \cite{wang2018rank1arxiv},
 and MVDR beamforming \cite{van1988beamforming},
 using the CHiME-3 data and a real internal test set.
These variants are compared
 with the state-of-the-art methods 
 of DNN-based beamforming\cite{heymann2016nngev}
 using phase-aware acoustic features 
 \cite{erdogan2016mask,pertila2017microphone,wang2018multi}
 and cost functions\cite{erdogan2015phase}.
In addition, we evaluate the performance 
 of the online extension of the proposed method.

The rest of this paper is organized as follows. 
Section~\ref{sec:related_work} describes related work on multichannel speech enhancement.
Section~\ref{sec:beamforming} and Section~\ref{sec:mnmf}
 explain three types of beamforming (full-rank MWF, rank-1 MWF, and MVDR beamforming)
 and MNMF, respectively.
Section~\ref{sec:proposed_method} explains the proposed method of MNMF-informed beamforming. 
Section~\ref{sec:evaluation} reports comparative experiments
 and Section~\ref{sec:conclusion} concludes the paper.

\section{Related Work}
\label{sec:related_work}

We review non-blind beamforming methods
 based on the steering vector of speech 
 and the SCM of noise
 for ASR in noisy environments.
We also review BSS methods including several variants of MNMF.

\subsection{Beamforming Methods}

There are several variants of beamforming
 such as DS \cite{anguera2007beamformit}
 MVDR \cite{van1988beamforming},
 GEV \cite{warsitz2007blind} beamforming
 and MWF \cite{souden2010optimal,wang2018rank1arxiv}.
The DS beamforming \cite{anguera2007beamformit} 
 uses only the steering vector of target speech
 and the other methods additionally use the SCM of noise. 
The GEV beamforming aims 
 to maximize the signal-to-noise ratio (SNR) \cite{warsitz2007blind}
 without putting any assumptions 
 on the acoustic transfer function from the speaker to the array 
 and the SCM of the noise.
The MVDR beamforming and the MWF, on the other hand,
 assume that the time-frequency (TF) bins of speech and noise spectrograms
 are distributed according to complex Gaussian distributions
 \cite{van1988beamforming,souden2010optimal,wang2018rank1arxiv}.
In Section~\ref{sec:beamforming},
 we review the relationships 
 between MVDR beamforming and rank-1 and full-rank MWF 
 in terms of the propagation process and and the filter estimation strategy.

TF Mask estimation has actively been studied
 for computing the SCMs of speech and noise
 \cite{higuchi2017online,nakatani2017integrating,heymann2016nngev,
 erdogan2016mask,xiao2017time,ochiai2017multichannel,pertila2017microphone,wang2018multi}.
Our unsupervised method is different from DNN-based mask estimation 
 \cite{nakatani2017integrating,heymann2016nngev,
 erdogan2016mask,xiao2017time,ochiai2017multichannel,pertila2017microphone,wang2018multi}
 in two ways.
First, our method decomposes each TF bin into the sum of speech and noise, 
 while the mask-based methods
 calculate the SCM of speech 
 from noisy TF bins without any decomposition.
Second, our method uses no training data, 
 while in general the DNN-based methods need
 a sufficient number of pairs of noisy spectrograms and ideal binary masks (IBMs).
The performance of the DNN-based mask estimation would be degraded 
 in unseen conditions that are not covered 
 by the training data because of overfitting to the training data.

The major limitation of most DNN-based methods 
 is that only single-channel magnitude spectrograms are used 
 for mask estimation
 by discarding the spatial information 
 such as ILDs and IPDs.
Recently, 
 Wang {\it et~al.}\cite{wang2018multi} and Pertil\"{a}\cite{pertila2017microphone}
 have investigated the use of ILDs and IPDs
 as acoustic features for mask estimation.
Erdogan {\it et~al.}\cite{erdogan2015phase}
 proposed a method for estimating a phase-sensitive filter
 in single-channel speech enhancement.
For comparative evaluation, inspired by these state-of-the-art methods,
 we use both spatial and magnitude features for DNN-based multichannel mask estimation.

\subsection{Multichannel Nonnegative Matrix Factorization}

Multichannel extensions of NMF
 \cite{itakura2018mnmf,ozerov2010mnmf,arberet2010mnmf,
 sawada2013mnmf,nikunen2014mnmf,kitamura2016mnmf}
 represent
 the complex spectrograms of multichannel mixture signals
 by using the SCMs and low-rank power spectrograms of multiple source signals.
Ozerov {\it et~al.}~\cite{ozerov2010mnmf}
 pioneered the use of NMF
 for multichannel source separation,
 where the SCMs are restricted to rank-1 matrices
 and the cost function based on the Itakura-Saito (IS) divergence is minimized.
This model was extended 
 to have full-rank SCMs~\cite{arberet2010mnmf}. 
Sawada {\it et~al.}~\cite{sawada2013mnmf}
 introduced partitioning parameters
 to have a set of basis spectra shared by all sources
 and derived a majorization-minimization (MM) algorithm.
Nikunen and Virtanen~\cite{nikunen2014mnmf}
 proposed a similar model
 that represents the SCM of each source
 as the weighted sum of direction-dependent SCMs.
While these methods can be used in a underdetermined case,
 Kitamura {\it et~al.}~\cite{kitamura2016mnmf}
 proposed independent low-rank matrix analysis (ILRMA) for a determined case
 by restricting the SCMs of \cite{sawada2013mnmf} to rank-1 matrices.
This can be viewed as a unified model of NMF and independent vector analysis (IVA)
 and is robust to initialization.

\section{Beamforming Methods}
\label{sec:beamforming}

This section introduces three major methods of beamforming;
 full-rank and rank-1 versions of multichannel Wiener filtering (MWF)
 and minimum variance distortionless response (MVDR) beamforming
 (Table~\ref{tab:beamforming})\cite{souden2010optimal}.

\begin{table}[t]
\centering
\caption{Beamforming methods.}
\label{tab:beamforming}
\vspace{-2mm}
 \begin{tabular}{c|c|c} 
 \toprule
 & Speech: Full-rank & Speech: Rank-1 \\
 & Noise: Full-rank & Noise: Full-rank \\
 \midrule
 MAP 
 & Full-rank MWF
 & Rank-1 MWF
 \\ 
 & Eqs.~(\ref{eq:w_wf_tv}) \& (\ref{eq:w_wf_ti})
 & Eqs.~(\ref{eq:w_wf1_tv}) \& (\ref{eq:w_wf1_ti}) 
 \\ 
 \midrule  
 ML 
 & - 
 & MVDR
 \\   
 & - 
 & Eqs. (\ref{eq:w_mv_tv}) \& (\ref{eq:w_mv_ti}) 
 \\ 
 \bottomrule
 \end{tabular}
\end{table}

\subsection{Overview}

The goal of beamforming is to 
 extract a source signal of interest from a mixture signal
 in the short-time Fourier transform (STFT) domain.
Let $\x_{ft}\in{\mathbb{C}}^{M}$ be the multichannel complex spectrum of the mixture
 at frequency $f$ and frame $t$
 recorded by $M$ microphones,
 which is assumed to be given by
\begin{align}
\x_{ft}
=
\x^\mathrm{s}_{ft} + \x^\mathrm{n}_{ft},
\label{eq:assumption_mwf}
\end{align}
where $\x^\mathrm{s}_{ft}$ and $\x^\mathrm{n}_{ft}$
 are the multichannel complex spectra of speech and noise (called {\it images}), respectively.
The notations are listed in Table~\ref{tab:notations}.
The goal is to estimate a linear demixing filter $\w_{ft}$ 
 that obtains an estimate $\hat{s}_{ft}$ of the target speech $s_{ft}$
 from the mixture (speech + noise) $\x_{ft}$ as follows:
\begin{align}
\hat{s}_{ft} = \w_{ft}^\hr \x_{ft}.
\label{eq:beamforming}
\end{align}

As shown in Table~\ref{tab:beamforming},
 the beamforming methods can be categorized
 in terms of sound propagation processes.
\begin{itemize}
\item
The {\it full-rank} propagation process
 considers various propagation paths caused by reflection and reverberation.
It is thus represented 
 by using an $M \times M$ full-rank SCM for each source.
\item
The {\it rank-1} propagation process considers only the direct paths
 from each sound source to the microphones.
 It is thus represented 
 by using an $M$-dimensional steering vector for each source.
\end{itemize}
The full-rank propagation process reduces to the rank-1 propagation process
 when the full-rank SCM is restricted to a rank-1 matrix
 whose eigenvector is equal to the steering vector.
 
The beamforming methods can also be categorized 
 in terms of estimation strategies.
\begin{itemize}
\item 
The {\it maximum a posteriori (MAP)} estimation
 assumes the target speech spectra to be complex Gaussian distributed.
\item
The {\it maximum likelihood (ML)} estimation
 uses no prior knowledge about the target speech spectra.
\end{itemize}

\begin{table}[t]
\centering
\caption{Mathematical notations.}
\label{tab:notations}
\vspace{-2mm}
 \begin{tabular}{c|ccc} 
  \toprule
  & {Observation} & {Speech} & {Noise} \\ 
  \midrule 
  Multichannel spectrum $\in \mathbb{C}^M$ & { $\x$ } & { $\x^\mathrm{s}$ } & { $\x^\mathrm{n}$ } \\ 
  Steering vector $\in \mathbb{C}^M$ & { - } & { $\p$ } & { $\q$ } \\  
  Spatial covariance matrix $\in \mathbb{S}_{++}^M$ & { $\X$ } & { $\P$ } & { $\Q$ } \\ 
  \bottomrule
 \end{tabular}
\end{table}

\subsection{Full-Rank Multichannel Wiener Filtering}
\label{sec:full_rank_mwf}

The full-rank MWF \cite{doclo2002gsvd}
 assumes both the target speech $\x^\mathrm{s}_{ft}$ and the noise $\x^\mathrm{n}_{ft}$
 to follow multivariate circularly-symmetric complex Gaussian distributions as follows:
\begin{align}
\x^\mathrm{s}_{ft} &\sim \ComplexGaussian(\zeros, \P_{ft}),
\label{eq:s_prior}
\\
\x^\mathrm{n}_{ft} &\sim \ComplexGaussian(\zeros, \Q_{ft}),
\label{eq:n_prior}
\end{align}
where $\P_{ft} \in \mathbb{S}_{++}^{M}$ and $\Q_{ft} \in \mathbb{S}_{++}^{M}$
 are the full-rank SCMs 
 of the speech and noise at frequency $f$ and time $t$, respectively,
 and $\mathbb{S}_{++}^{M}$ indicates
 the set of $M \times M$ Hermitian positive definite matrices.
Using the reproducible property of the Gaussian distribution,
 we have
\begin{align}
 \x_{ft} &\sim \ComplexGaussian(\zeros, \P_{ft} + \Q_{ft})\textcolor{red}{.}
\end{align}

Given the mixture $\x_{ft}$,
 the posterior distribution 
 of the multichannel speech \textit{image} $\x^\mathrm{s}_{ft}$ is obtained as follows:
\begin{align}
 \x^\mathrm{s}_{ft} \mid \x_{ft} 
 &\sim 
 \ComplexGaussian
 \!\left(
 \P_{ft} \left(\P_{ft} + \Q_{ft}\right)^{-1} \x_{ft}, \right.
 \\
 &\ \ \ \ \ \ \ \ \
 \left.
 \P_{ft} - \P_{ft} \left(\P_{ft} + \Q_{ft}\right)^{-1} \P_{ft}
 \right).
\end{align}
To obtain a monaural estimate $\hat{s}_{ft}$ of the speech,
 it is necessary to choose a reference channel (dimension) $m$
 from the MAP estimate of the speech image $\x^\mathrm{s}_{ft}$.
The time-variant demixing filter $\w_{ft}^\wf$ is thus given by
\begin{align}
\w_{ft}^\wf(m)
=
({\P_{ft}}+{\Q_{ft}})^{-1} {\P_{ft}} \u_m,
\label{eq:w_wf_tv}
\end{align}
 where $\u_m \in{\mathbb{C}}^{M}$ is the $M$-dimensional one-hot vector
 that takes $1$ in dimension $m$.
If the speaker does not move and the noise is stationary,
 $\P_{ft}$ and $\Q_{ft}$
 are often assumed to be time-invariant, {\it i.e.},
 $\P_{ft} = \P_{f}$ and $\Q_{ft} = \Q_{f}$.
In this case, the time-invariant demixing filter $\w_{f}^\wf$
 is given by
\begin{align}
\w_{f}^\wf(m)
=
({\P_{f}}+{\Q_{f}})^{-1} {\P_{f}} \u_m.
\label{eq:w_wf_ti}
\end{align}
In reality, the speech is not stationary,
 but such time-invariant linear filtering is known to be effective 
 for speech enhancement with small distortion.
In general, the enhanced speech signals obtained by the time-variant filter
 tends to be more distorted than those obtained by the time-invariant filter.
 
The optimal reference channel $m^*$ is chosen such that
 the average-a-posteriori SNR is maximized with respect to $m$
 as follows \cite{erdogan2016mask,lawin2012reference}:
\begin{align}
m^*
=
\Argmax_{m}
\frac
{\sum_{t=1}^{T} \sum_{f=1}^{F} \w_{ft}^\hr(m) \P_{ft} \w_{ft}(m)}
{\sum_{t=1}^{T} \sum_{f=1}^{F} \w_{ft}^\hr(m) \Q_{ft} \w_{ft}(m)},
\label{eq:optimal_m}
\end{align}
where $\w_{ft}(m)$ is a demixing filter
 obtained by assuming the reference channel to be $m$.
Different microphones may thus be selected for individual utterances.

\subsection{Rank-1 Multichannel Wiener Filtering}
\label{sec:rank_one_mwf}

The rank-1 MWF \cite{wang2018rank} is obtained
 as a special case of the full-rank MWF
 when the spatial covariance matrix $\P_{ft}$ of the speech
 is restricted to a rank-1 matrix as follows:
\begin{align}
 \P_{ft} = \lambda_{ft} \p_{ft} \p_{ft}^\hr,
 \label{eq:wm1_constraint}
\end{align}
 where $\lambda_{ft}$ and $\p_{ft} \in \mathbb{C}^M$
 are the power and steering vector of the speech
 at frequency $f$ and time $t$, respectively.
Substituting Eq.~(\ref{eq:wm1_constraint}) into Eq.~(\ref{eq:w_wf_tv})
 and using the Woodbury matrix identity,
 we obtain the time-variant demixing filter $\w_{ft}^\wfone$ as follows:
\begin{align}
 \w_{ft}^\wfone(m)
 =
 \frac{\Q_{ft}^{-1} \p_{ft}}
 {\p_{ft}^\hr \Q_{ft}^{-1} \p_{ft} + \lambda_{ft}^{-1}}
 \p_{ft}^\hr \u_m.
 \label{eq:w_wf1_tv}
\end{align}
In practice, to achieve reasonable performance,
 we assume the time-invariance of the speech,
 {\it i.e.}, $\lambda_{ft} = \lambda_{f}$ and $\p_{ft} = \p_{f}$.
Similarly, the time-invariant filter $\w_{f}^\wfone$ is given by
\begin{align}
 \w_{f}^\wfone(m)
 =
 \frac{\Q_{f}^{-1} \p_{f}}
 {\p_{f}^\hr \Q_{f}^{-1} \p_{f} + \lambda_{f}^{-1}}
 \p_{f}^\hr \u_m.
 \label{eq:w_wf1_ti}
\end{align}
Given the steering vector $\p_{ft}$ or $\p_{f}$,
 the power spectral density
 $\lambda_{ft}$ in Eqs.~(\ref{eq:w_wf1_tv}) or $\lambda_{f}$ in (\ref{eq:w_wf1_ti}),
 can be estimated as follows:
\begin{align}
 \lambda_{ft}
 \simeq
 \frac{\| \P_{ft} \|}{\| \p_{ft} \p_{ft}^\hr \|},
 \ \ \
 \lambda_{f}
 \simeq
 \frac{\|\P_{f}\|}{\|\p_{f}\p_{f}^\hr\|},
\end{align}
where $\|\cdot\|$ represents the Frobenius norm of a matrix.

\subsection{Minimum Variance Distortionless Response Beamforming}

The MVDR beamforming \cite{van1988beamforming}
 can be derived as a special case of the rank-1 MWF
 when the power spectral density of the speech in Eq.~(\ref{eq:wm1_constraint})
 (the variance of the Gaussian distribution in Eq.~(\ref{eq:s_prior})) approaches infinity,
 {\it i.e.}, we do not put any assumption on the target speech.
The time-variant and time-invariant demixing filters are given by
\begin{align}
 \w_{ft}^\mv(m)
 &=
 \frac{\Q_{ft}^{-1} \p_{ft}}
 {\p_{ft}^\hr \Q_{ft}^{-1} \p_{ft}}
 \p_{ft}^\hr \u_m,
 \label{eq:w_mv_tv}
 \\
 \w_{f}^\mv(m)
 &=
 \frac{\Q_{f}^{-1} \p_{f}}
 {\p_{f}^\hr \Q_{f}^{-1} \p_{f}}
 \p_{f}^\hr \u_m.
 \label{eq:w_mv_ti} 
\end{align}

\section{Multichannel Nonnegative Matrix Factorization}
\label{sec:mnmf}

This section introduces 
 multichannel nonnegative matrix factorization (MNMF)\cite{sawada2013mnmf}.
In this paper 
 we assume that the observed noisy speech contains $N$ sound sources,
 one of which corresponds to target speech
 and the other sources are regarded as noise.
Let $M$ be the number of channels (microphones).

\subsection{Probabilistic Formulation}
\label{sec:probabilistic_formulation}

We explain the generative process of the multichannel observations of noisy speech, $\X = \{\x_{ft}\}_{f=1,t=1}^{F,T}$,
 where $\x_{ft} \in \mathbb{C}^M$ is the multichannel complex spectrum of the mixture
 at frequency $f$ and time $t$.
Let $s_{nft} \in \mathbb{C}$ be the single-channel complex spectrum of source $n$ 
 at frequency $f$ and time $t$
 and $\x_{nft} \in \mathbb{C}^M$ be the multichannel complex spectrum (image) of source $n$.
If the sources do not move, we have
\begin{align}
 \x_{nft} = \g_{nf} s_{nft},
 \label{eq:x_nft}
\end{align}
where $\g_{nf}\in{\mathbb{C}}^{M}$ is 
 the time-invariant steering vector of source $n$ at frequency $f$.
Here $s_{ft}$ is assumed to be circularly-symmetric complex Gaussian distributed as follows:
\begin{align}
 s_{nft} \sim \ComplexGaussian\!\left(0, \lambda_{nft} \right),
 \label{eq:s_nft_dist}
\end{align}
where $\lambda_{nft} \ge 0$ is the power spectral density 
 of source $n$ at frequency $f$ and time $t$.
Using Eq.~(\ref{eq:x_nft}) and Eq.~(\ref{eq:s_nft_dist}),
 $\x_{nft}$ can be said to be 
 multivariate circularly-symmetric complex Gaussian distributed as follows:
\begin{align}
 \x_{nft} \sim \ComplexGaussian\!\left(\zeros, \lambda_{nft} \G_{nf}\right),
 \label{eq:x_nft_dist}
\end{align}
where 
 and $\G_{nf} = \g_{nf} \g_{nf}^\hr$ 
 is the rank-1 SCM of source $n$ at frequency $f$.
In MNMF, the rank-1 assumption on $\G_{nf}$
 is relaxed to deal with the underdetermined condition of $M < N$
 by allowing  $\G_{nf} \in \mathbb{S}_{++}^{M}$ 
 to take any full-rank positive definite matrix.
Assuming the instantaneous mixing process (source additivity) in the frequency domain,
 we have
\begin{align}
 \x_{ft} = \sum_{n=1}^N \x_{nft}.
 \label{eq:x_ft}
\end{align}
Using Eq.~(\ref{eq:x_nft_dist}) and Eq.~(\ref{eq:x_ft}), 
 the reproducible property of the Gaussian distribution leads to
\begin{align}
 \x_{ft} \sim \ComplexGaussian\!\left(\zeros, \sum_{n=1}^N \lambda_{nft} \G_{nf}\right).
 \label{eq:x_ft_gauss} 
\end{align}

The nonnegative power spectral density $\lambda_{nft}$ of each source $n$
 is assumed to be factorized in an NMF style as follows:
\begin{align}
 \lambda_{nft} = \sum_{k=1}^K v_{nkf} h_{nkt},
\end{align}
where $K$ is the number of basis spectra,
 $v_{nkf} \ge 0$ is the power of basis $k$ at frequency $f$
 and $h_{nkt} \ge 0$ is the activation of basis $k$ at time $t$.
This naive model has $NK$ basis spectra in total.
One possibility to reduce the number of parameters
 is to share $K$ basis spectra between all $N$ sources as follows:
\begin{align}
 \lambda_{nft} = \sum_{k=1}^K z_{nk} v_{kf} h_{kt},
 \label{eq:lambda_fact}
\end{align}
where $z_{nk}$ indicates the weight of basis $k$ in source $n$.
Substituting Eq.~(\ref{eq:lambda_fact}) into Eq.~(\ref{eq:x_ft_gauss}),
 we obtain the probabilistic generative model of $\X$ as follows:
\begin{align}
\x_{ft} \sim \ComplexGaussian\!\left(\zeros, \sum_{k=1}^K v_{kf} h_{kt} \sum_{n=1}^N z_{nk} \G_{nf}\right).
\label{eq:model_MNMF}
\end{align}

\subsection{Parameter Estimation}

Given $\X$,
 our goal is to estimate $\V = \{v_{kf}\}_{k=1,f=1}^{K,F}$, 
$\H = \{h_{kt}\}_{k=1,t=1}^{K,T}$, 
$\Z = \{z_{nk}\}_{n=1,k=1}^{N,K}$, 
and $\G = \{\G_{nf}\}_{n=1,f=1}^{N,F}$
 that maximize the likelihood obtained 
 by multiplying Eq.~(\ref{eq:model_MNMF}) over all frequency $f$ and time $t$.
Let 
 two positive definite matrices $\X_{ft} \in \mathbb{S}_{++}^{M}$ 
 and $\Y_{ft} \in \mathbb{S}_{++}^{M}$ be as follows:
\begin{align}
\X_{ft} &= \x_{ft} \x_{ft}^\hr,
\\
\Y_{ft} &= \sum_{k=1}^K v_{kf} h_{kt} \sum_{n=1}^N z_{nk} \G_{nf}.
\end{align}
The maximization of the likelihood function given by Eq.~(\ref{eq:model_MNMF})
 is equivalent to the minimization of the log-determinant divergence 
 between $\X_{ft}$ and $\Y_{ft}$ given by
\begin{align}
\!\!\!
\mathcal{D}_{\scr{LD}}(\X_{ft}|\Y_{ft})
=
\mbox{tr}\!\left(\X_{ft} \Y_{ft}^{-1}\right)
- \log\!\left|\X_{ft} \Y_{ft}^{-1}\right| - M,
\end{align}
The total cost function $f(\V, \H, \Z, \G)$ to be minimized 
w.r.t.\@ $\V$, $\H$, $\Z$, and $\G$ is thus given by
\begin{align}
f(\V, \H, \Z, \G)
=
\sum_{f=1}^{F} \sum_{t=1}^{T} \mathcal{D}_{\scr{LD}}(\X_{ft}|\Y_{ft}).
\label{eq:distance_MNMF}
\end{align}

Since Eq.~(\ref{eq:distance_MNMF}) is hard to directly minimize,
 a convergence-guaranteed MM algorithm was proposed 
 (see \cite{sawada2013mnmf} for detailed derivation).
The updating rules are given by
\begin{align}
v_{kf}
&\gets
v_{kf}
\sqrt{
\frac
{\sum_{n} {z}_{nk} \sum_{t} h_{kt} {\rm tr}\!\left(\Y_{ft}^{-1} \X_{ft} \Y_{ft}^{-1} \G_{nf} \right)}
{\sum_{n} {z}_{nk} \sum_{t} h_{kt} {\rm tr}\!\left( \Y_{ft}^{-1} \G_{nf} \right)}
},
\label{eq:mnmf_offline_b}
\\
h_{kt}
&\gets
h_{kt}
\sqrt{
\frac
{\sum_{n} {z}_{nk} \sum_{f} v_{kf} {\rm tr}\!\left(\Y_{ft}^{-1} \X_{ft} \Y_{ft}^{-1} \G_{nf} \right)}
{\sum_{n} {z}_{nk} \sum_{f} v_{kf} {\rm tr}\!\left( \Y_{ft}^{-1} \G_{nf} \right)}
},
\label{eq:mnmf_offline_c}
\\
{z}_{nk}
&\gets
{z}_{nk}
\sqrt{
\frac
{\sum_{f} v_{kf} \sum_{t} h_{kt} {\rm tr}\!\left(\Y_{ft}^{-1} \X_{ft} \Y_{ft}^{-1} \G_{nf} \right)}
{\sum_{f} v_{kf} \sum_{t} h_{kt} {\rm tr}\!\left(\Y_{ft}^{-1} \G_{nf} \right)}
}.
\label{eq:mnmf_offline_z}
\end{align}
$\G_{nf}$ is obtained as the unique solution of a special case 
 of the continuous time algebraic Riccati equation
 $\G_{nf} \mPsi_{nf} \G_{nf} 
 = \G_{nf}^{\scr{old}} \mPhi_{nf} \G_{nf}^{\scr{old}}$.
In the original study on MNMF \cite{sawada2013mnmf},
 this equation was solved using an iterative optimization algorithm.
In the field of information geometry, however,
 the analytical solution of this equation is known to exist as follows:
\begin{align}
\mPhi_{nf}
&=
\sum_{k}{z}_{nk}v_{kf} \sum_{t}h_{kt}\Y_{ft}^{-1}\X_{ft}\Y_{ft}^{-1},
\label{eq:mnmf_offline_H2}
\\
\mPsi_{nf}
&=
\sum_{k}{z}_{nk} v_{kf} \sum_{t}h_{kt}\Y_{ft}^{-1},
\label{eq:mnmf_offline_H1}
\\
\G_{nf}
&\gets
\mPsi_{nf}^{-\frac{1}{2}} 
\left(\mPsi_{nf}^{\frac{1}{2}} \G_{nf} \mPhi_{nf} \G_{nf} \mPsi_{nf}^{\frac{1}{2}} \right)^{\frac{1}{2}} 
\mPsi_{nf}^{-\frac{1}{2}},
\label{eq:mnmf_offline_H}
\end{align}
where $\G_{nf}$ is updated to
 the geometric mean of two positive definite matrices
 $\mPsi_{nf}^{-1}$ and $\G_{nf}^{\scr{old}} \mPhi_{nf} \G_{nf}^{\scr{old}}$
 \cite{ando2004geometric,chen2015review,yoshii2018icassp}.

\section{MNMF-Informed Beamforming}
\label{sec:proposed_method}

This section explains the proposed MNMF-informed beamforming and its online extension.
Our method takes as input the multichannel noisy speech spectrograms $\X$
 and outputs a speech spectrogram, 
 which is then passed to an ASR back-end (Fig.~\ref{fig:pm_framework_intro}).
MNMF is used to estimate 
 the SCMs of speech and the other sounds from $\X$.
The steering vector $\p$ of the target speech 
 and the SCM $\Q$ of noise are then computed.
Finally, the enhanced speech is obtained by using 
 one of the three kinds of beamforming
 described in Section~\ref{sec:beamforming}.

\subsection{Estimation of Spatial Information}
\label{sec:estimation}

To use a beamforming method (Section~\ref{sec:beamforming}), 
 we compute the SCMs $\P$ and $\Q$ of speech and noise
 by using the parameters $\V$, $\H$, $\Z$, and $\G$ of MNMF 
 (Section~\ref{sec:mnmf}).
Assuming that source $n=1$ is the target speech (see Section~\ref{sec:initialization}),
 we have 
\begin{align}
\P_{ft}
&=
\sum_{k=1}^{K} v_{kf} h_{kt} z_{1k} \G_{f1},
\label{eq:P_ft}
\\
\Q_{ft}
&=
\sum_{k=1}^{K} v_{kf} h_{kt} \sum_{n=2}^{N} z_{nk} \G_{nf},
\label{eq:Q_ft}
\end{align}
where $\P_{ft}$ and $\Q_{ft}$ are
 the time-variant SCMs of speech and noise, respectively.
The time-invariant SCMs $\P_{f}$ and $\Q_{f}$ are also given by
\begin{align}
\P_{f}
&=
\frac{1}{T}\sum_{t=1}^{T}\P_{ft},
\label{eq:P_f}
\\
\Q_{f}
&=
\frac{1}{T}\sum_{t=1}^{T}\Q_{ft}.
\label{eq:Q_f}
\end{align}
The corresponding steering vectors $\p_{ft}$ and $\p_{f}$ of the target speech 
 are approximated as the principal components of $\P_{ft}$ and $\P_{f}$, respectively,
 as follows:
\begin{align}
\p_{ft}
&=
\mathcal{PE}\!\left(\P_{ft}\right),
\label{eq:pe_tv}
\\
\p_{f}
&=
\mathcal{PE}\!\left(\P_{f}\right).
\label{eq:pe_ti}
\end{align}

\subsection{Online MNMF}
\label{sec:online}

We propose an online extension of MNMF
 that incrementally updates the parameters $\V$, $\H$, $\Z$, and $\G$.
Suppose that $\X$ is given as a series of $J$ mini-batches
 in a sequential order, where
 each mini-batch $j$ consists of multiple frames $\in {t}^{(j)}$.
The notation $*^{(j)}$ represents a statistic of mini-batch $j$.
The latest statistics are considered with a weight $\rho$.
When $\rho < 1$, the current mini-batch is put more emphasis \cite{lefevre2011online}.
The online updating rules are as follows:
\begin{align}
\begin{split}
{\alpha}_{kf}^{(j)}
&=
{\sum_{n}{z}_{nk}^{(j)} \sum_{t \in {t}^{(j)} }h_{kt} 
{\rm tr}\!\left(\Y_{ft}^{-1} \X_{ft} \Y_{ft}^{-1} \G_{nf}^{(j)} \right)},
\label{eq:mnmf_online_b1}
\end{split}
\\
{\beta}_{kf}^{(j)}
&=
\sum_{n}{z}_{nk}^{(j)} \sum_{t \in {t}^{(j)} }h_{kt} 
{\rm tr}\!\left( \Y_{ft}^{-1} \G_{nf}^{(j)} \right),
\label{eq:mnmf_online_b2}
\\
v_{kf}^{(j)}
&\gets
\sqrt{
\frac 
{F_j(v_{kf}, {\alpha}_{kf}, v_{kf})}
{F_j({\beta}_{kf})}
},
\label{eq:mnmf_online_b}
\\
\begin{split}
{\gamma}_{nk}^{(j)}
&=
{\sum_{f}v_{kf}^{(j)} \sum_{t \in {t}^{(j)} }h_{kt}
{\rm tr}\!\left(\Y_{ft}^{-1} \X_{ft} \Y_{ft}^{-1} \G_{nf}^{(j)} \right)},
\label{eq:mnmf_online_z1}
\end{split}
\\
{\delta}_{nk}^{(j)}
&=
\sum_{f}v_{kf}^{(j)} \sum_{t \in {t}^{(j)} }h_{kt} {\rm tr}\!\left(\Y_{ft}^{-1} \G_{nf}^{(j)} \right),
\label{eq:mnmf_online_z2}
\\
{z}_{nk}^{(j)}
&\gets
\sqrt{
\frac
{F_j(z_{nk}, {\gamma}_{nk}, z_{nk})}
{F_j({\delta}_{nk})}
},
\label{eq:mnmf_online_z}
\\
\begin{split}
\mPhi_{nf}^{(j)}
&=
\sum_{k}{z}_{nk}^{(j)}v_{kf}^{(j)} \sum_{t \in {t}^{(j)} }h_{kt}\Y_{ft}^{-1}\X_{ft}\Y_{ft}^{-1},
\label{eq:mnmf_online_H2}
\end{split}
\\
\mPsi_{nf}^{(j)}
&=
\sum_{k}{z}_{nk}^{(j)} v_{kf}^{(j)} \sum_{t \in {t}^{(j)} }h_{kt}\Y_{ft}^{-1},
\label{eq:mnmf_online_H1}
\\
\G_{nf}^{(j)}
&\gets
F_j(\mPsi_{nf})^{-\frac{1}{2}}
\nonumber\\
&\ \ \
\left(
F_j^{\frac{1}{2}} (\mPsi_{nf})
F_j(\G_{nf}, \mPhi_{nf}, \G_{nf}) 
F_j^{\frac{1}{2}} (\mPsi_{nf})
\right)^{\frac{1}{2}} 
\nonumber\\
&\ \ \
F_j(\mPsi_{nf})^{-\frac{1}{2}},
\label{eq:mnmf_online_H}
\end{align}
where the function $F_j$ is defined as follows:
\begin{align}
F_j(x)
&= x^{(j)} + \rho x^{(j - 1)},
\\
F_j(a, x, a)
&= a^{(j)} x^{(j)} a^{(j)} + \rho a^{(j-1)} x^{(j-1)} a^{(j-1)}.
\end{align}

\subsection{Initialization of MNMF}
\label{sec:initialization}

We randomly initialize all parameters except for $\G$.
Since MNMF is sensitive to the initialization of $\G$ \cite{tachioka2017coupled},
 we use a constrained version of MNMF called 
 independent low-rank matrix analysis (ILRMA) \cite{kitamura2016mnmf}
 for initializing $\G$.
Since ILRMA can be used in the determined condition of $M=N$,
 in this paper we assume $M=N$ for MNMF.
In ILRMA, 
 $\G_{nf}$ is restricted to a rank-1 matrix,
 {\it i.e.}, $\G_{nf} = \g_{nf} \g_{nf}$ 
 (see Section \ref{sec:probabilistic_formulation}).
Using Eq.~(\ref{eq:x_nft}) and Eq.~(\ref{eq:x_ft}),
 we have
\begin{align}
 \x_{ft} = \G_f \s_{ft},
\end{align}
where $\s_{ft} = [s_{1ft},\cdots,s_{Nft}]^\tr \in \mathbb{C}^{N}$ 
 is a set of source spectra 
 and  $\G_f = [\g_{1f},\cdots,\g_{Nf}] \in \mathbb{C}^{M \times N}$ 
 is a mixing matrix.
If $\G_f$ is a non-singular matrix, we have
\begin{align}
 \s_{ft} = \W_f^\hr \x_{ft},
\end{align}
where $\W_f^\hr = [\w_{1f},\cdots,\w_{Nf}]^\hr = \G_f^{-1}$ is a demixing matrix
 and $\w_{nf} \in \mathbb{C}^M$ is a demixing filter of source $n$.
We use ILRMA for estimating $\W_f$,
 compute $\G_f = \W_f^{-\hr}$,
 and initialize $\G_{nf} = \g_{nf} \g_{nf} + \epsilon \I$,
 where $\epsilon$ is a small number.

\begin{algorithm}[t]
\caption{Offline MNMF-informed beamforming.}
\begin{algorithmic}[1]
\STATE{Initialize $\G$ by ILRMA}
\FOR{iteration $= 1$ to MaxIteration }
\STATE{Update $\V$ by Eq.~(\ref{eq:mnmf_offline_b})}
\STATE{Update $\H$ by Eq.~(\ref{eq:mnmf_offline_c})}
\STATE{Update $\Z$ by Eq.~(\ref{eq:mnmf_offline_z})}
\STATE{Update $\G$ by Eq.~(\ref{eq:mnmf_offline_H})}
\ENDFOR
\STATE{Estimate $\P$ by Eq.~(\ref{eq:P_ft}) or (\ref{eq:P_f})}
\STATE{Estimate $\Q$ by Eq.~(\ref{eq:Q_ft}) or (\ref{eq:Q_f})}
\STATE{Estimate $\p$ by Eq.~(\ref{eq:pe_tv}) or (\ref{eq:pe_ti})}
\STATE{Estimate $\w$ by Eq.~(\ref{eq:w_wf_tv}), (\ref{eq:w_wf_ti}),
(\ref{eq:w_wf1_tv}), (\ref{eq:w_wf1_ti}),
(\ref{eq:w_mv_tv}), or (\ref{eq:w_mv_ti})}
\STATE{Estimate ${y}$ by Eq.~(\ref{eq:beamforming})}
\end{algorithmic}
\label{alg:offline}
\end{algorithm}

In this paper,
 we assume that the target speech is predominant 
 in the duration of $\X$ ({\it e.g.}, one utterance).
To deal with longer observations,
 voice activity detection (VAD) would be needed
 for segmenting the signals into multiple utterances.
In reality, it can be said to be rare that 
 a target utterance largely overlaps another utterance with the same level of volume.
To make source $1$ correspond to the target speech,
 the steering vector $\g_{f1}$ of source $1$ is thus initialized 
 as the principal component of the average empirical SCM as follows:
\begin{align}
\g_{f1}^{\rm init}
=
\mathcal{PE}\left(\frac{1}{T} \sum_{t=1}^T \X_{ft}\right).
\label{eq:principal_eigenvector_initialization}
\end{align}
In the online version,
 the average of the empirical SCMs is taken over the first mini-batch.

The procedures of the offline and online versions of speech enhancement 
 are shown in Algorithm~\ref{alg:offline} and Algorithm~\ref{alg:online}.
In the online version,
 the spatial information of the target speech and noise
 are initialized by using the first relatively-long mini-batch ({\it e.g.}, 10 s),
 and then updated in each mini-batch ({\it e.g.}, 0.5 s).
As described in Section~\ref{sec:rank_one_mwf},
 when the time-variant rank-1 MWF is used,
 the SCM $\P_{ft}$, the steering vector $\p_{ft}$, and the power $\lambda_{ft}$ 
 of the speech are assumed to be time-invariant,
 while those of noise are kept to be time-variant.

\begin{algorithm}[t]
\caption{Online MNMF-informed beamforming.}
\begin{algorithmic}[1]
\STATE{Initialize $\G^{(1)}$ by ILRMA}
\FOR{$j = 1$ to $J$}
\FOR{iteration $= 1$ to MaxIteration}
\STATE{Update $\V^{(j)}$ by Eq.~(\ref{eq:mnmf_online_b})}
\STATE{Update $\H^{(j)}$ by Eq.~(\ref{eq:mnmf_offline_c})}
\STATE{Update $\Z^{(j)}$ by Eq.~(\ref{eq:mnmf_online_z})}
\STATE{Update $\G^{(j)}$ by Eq.~(\ref{eq:mnmf_online_H})}
\ENDFOR
\STATE{Estimate $\P^{(j)}$ by Eq.~(\ref{eq:P_ft}) or (\ref{eq:P_f})}
\STATE{Estimate $\Q^{(j)}$ by Eq.~(\ref{eq:Q_ft}) or (\ref{eq:Q_f})}
\STATE{Estimate $\p^{(j)}$ by Eq.~(\ref{eq:pe_tv}) or (\ref{eq:pe_ti})}
\STATE{Estimate $\w^{(j)}$ by Eq.~(\ref{eq:w_wf_tv}), (\ref{eq:w_wf_ti}),
(\ref{eq:w_wf1_tv}), (\ref{eq:w_wf1_ti}), 
(\ref{eq:w_mv_tv}), or (\ref{eq:w_mv_ti})}
\STATE{Estimate ${y}^{(j)}$ by Eq.~(\ref{eq:beamforming})}
\ENDFOR
\end{algorithmic}
\label{alg:online}
\end{algorithm}

\section{Evaluation}
\label{sec:evaluation}

This section reports comprehensive experiments conducted 
 for evaluating all the variants of the proposed method
 based on unsupervised MNMF-informed beamforming 
 ({\it i.e.}, full-rank MWF, rank-1 MWF, or MVDR,
 time-variant or time-invariant, and offline or online),
 in comparison with the state-of-the-art methods 
 based on supervised DNN-based mask estimation.
To evaluate the performance of ASR,
 we used a common dataset
 taken from the third CHiME Challenge \cite{barker2015chime}, 
 where a sufficient amount of training data are available.
We also used an internal dataset consisting 
 of multichannel recordings in real noisy environments
 whose acoustic characteristics were different 
 from those of the ChiME-3 dataset.

\subsection{Configurations}

We describe the configurations 
 of the speech enhancement methods used for evaluation.
The multichannel complex spectrograms 
 of noisy speech signals recorded 
 by five or six microphones at a sampling rate of 16 [kHz]
 were obtained by short-time Fourier transform (STFT)
 with a Hamming window of 1024 samples (160 [ms]) 
 and a shifting interval of 160 samples (10 [ms]),
 {\it i.e.,} $M=5 \ \mathrm{or} \ 6$ and $F=513$.

\subsubsection{MNMF-Informed Beamforming}

In MNMF,
 the number of basis spectra was set as $K=25$
 and the number of sources was set as $N=5$
 (one source for speech and the remaining four sources for noise).
MNMF was combined with the time-variant and time-invariant versions
 of full-rank MWF, rank-1 MWF, 
 and MVDR beamforming (MNMF-\{TV, TI\}-\{WF, WF1, MV\}).
The demixing filter
 was computed from the same SCMs estimated by MNMF
 to prevent the initialization sensitivity
 from affecting the ASR performance.

\subsubsection{DNN-Based Beamforming}

For comparison,
 we used the time-invariant version of MVDR beamforming based on DNN-based mask estimation
 because the time-invariant version of MVDR beamforming
 has been the most common choice among various kinds of beamforming
 in DNN-based beamforming.

To estimate masks,
 we used different combinations of magnitude and spatial features \cite{wang2018multi}
 described below:
\begin{itemize}
\item 
The log of the outputs of $D$-channel mel-scale filter banks (LMFBs)
 were computed at each time $t$
 from the magnitude spectrogram of 
 a reference channel $m^*$ manually specified 
 or automatically selected by Eq.~\eqref{eq:optimal_m},
 where we set $D=100$.
These features were stacked 
 over 11 frames from time $t - 5$ to time $t + 5$ at each time $t$.
\item 
The $(M-1)$-dimensional ILDs and IPDs 
 (sine and cosine of phase angle differences)
 from the reference channel $m^*$
 were extracted at each frame $t$ and frequency $f$.
This is considered to be more robust to over-fitting
 than using all the $\binom{M}{2}$-dimensional ILDs and IPDs 
 between $M$ channels as proposed in \cite{wang2018multi}.
\end{itemize}
These features were stacked 
 over 11 frames from time $t - 5$ to time $t + 5$
 to obtain up to $(11D + 3F(M - 1))$-dimensional features at each time $t$,
 which were fed into DNNs.

To train DNNs, we tested two kinds of cost functions 
 with different target data \cite{erdogan2015phase} described below:
\begin{itemize}
\item
Ideal binary masks (IBMs) are used as target data,
 {\it i.e.}, a TF mask $a_{ft}$ takes $1$ 
 when $|x^\mathrm{s}_{ftm^*}| > |x^\mathrm{n}_{ftm^*}|$
 and takes $0$ otherwise,
 as in the standard DNN-based mask estimation.
The cost function is based on the cross-entropy loss 
 between the target masks and the outputs of a DNN.
\item
Phase-sensitive filters (PSFs) are used as target data,
 {\it i.e.}, a TF {\it filter} $a_{ft}$ is defined 
 as $a_{ft} = \mathrm{Real}(x^\mathrm{s}_{ftm^*} / x_{ftm^*})$,
 as proposed in \cite{erdogan2015phase}.
The cost function is based on the phase-sensitive spectrum approximation (PSA)
 between the filtered and ground-truth speech spectra
 given by $\mathcal{D}_{\scr{PSA}} = |a_{ft} x_{ftm^*} - x^\mathrm{s}_{ftm^*}|^2$.
\end{itemize}

We defined a baseline using LMFBs and IBMs (DNN-IBM)
 and its counterpart using LMFBs and PSFs (DNN-PSF).
As extensions of DNN-IBM,
 we tested the additional use of ILDs and/or IPDs
 (DNN-IBM-\{L, P, LP\}).
A standard feed-forward DNN was trained under each configuration.
Although a bidirectional long short-term memory network (BLSTM) 
 was originally proposed for DNN-IBM \cite{heymann2016nngev},
 the feed-forward DNN slightly outperformed the BLSTM in our preliminary experiments.
We thus report the results obtained with the feed-forward DNN in this paper.
The steering vector $\p_f$ of speech and the SCM of noise $\Q_f$
 used in Eq.~(\ref{eq:w_mv_ti}) 
 are given by
\begin{align}
\p_{f}
&=
\mathcal{PE}\left(\frac{1}{\sum_{t=1}^{T} a_{ft}} \sum_{t=1}^{T} a_{ft} \X_{ft}\right),
\label{eq:P_ft_dnn}
\\
\Q_{f}
&=
\frac{1}{\sum_{t=1}^{T} 1 - a_{ft}} \sum_{t=1}^{T} (1 - a_{ft}) \X_{ft}.
\label{eq:Q_ft_dnn}
\end{align}
As a common baseline,
 the weighted delay-sum (DS) beamforming called {\it Beamformit}
 \cite{anguera2007beamformit} was also used for comparison.

\subsubsection{Automatic Speech Recognition}

We used a de-facto standard ASR system based on
 a DNN-HMM \cite{mohamed2012acoustic,dahl2012context}
 and a standard WSJ 5k trigram model as acoustic and language models, respectively,
 with the Kaldi WFST decoder \cite{povey2011kaldi}.
The DNN had four hidden layers 
 with 2,000 rectified linear units (ReLUs) \cite{nair2010rectified} 
 and a softmax output layer with 2,000 nodes.
Its input was a 1,320-dimensional feature vector 
 consisting of 11 frames of 40-channel LMFB outputs 
 and their delta and acceleration coefficients.
Mean and variance normalization was applied to input vectors.
Dropout \cite{srivastava2014dropout} and batch normalization \cite{ioffe2015batch} 
 were used in the training of all hidden layers.

\subsubsection{Performance Evaluation}

The performance of ASR was measured in terms of the word error rate (WER)
 defined as the ratio of the number of substitution, deletion, and insertion errors 
 to the number of words in the reference text.
The performance of speech enhancement 
 was measured in terms of the speech distortion ratio (SDR) \cite{vincent2006performance}
 defied as the ratio of the energy of target components
 to that of distortion components 
 including interference, noise, and artifact errors.
In addition,
 the performance of speech enhancement 
 was measured in terms of 
 the perceptual evaluation of speech quality (PESQ) \cite{rix2001pesq}
 and short-time objective intelligibility (STOI) \cite{taal2011stoi}
 which are closely related to the human auditory perception.

\begin{table*}[t]
 \centering
 \caption{The experimental results of noisy speech recognition (WERs) 
 for the simulated and real evaluation data of CHiME-3.}
 \label{tab:chime_wer_results}
 \vspace{-2mm}
 \begin{tabular}{l|l|cc|ccccc|ccccc}
 \toprule
 & SCM estimation
 & \multicolumn{2}{c|}{Beamforming} 
 & \multicolumn{5}{c|}{Simulated data} 
 & \multicolumn{5}{c}{Real data}\\
 Method & (Target / Features) & Time & Type
 & BUS & CAF & PED & STR & Av.  
 & BUS & CAF & PED & STR & Av. \\ 
 \midrule 
 Not enhanced & & &
 & 11.64 & 17.18 & 14.05 & 15.33 & 14.55 & 31.00 & 24.62 & 18.33 & 14.81 & 22.19 \\ 
 Beamformit   & & Inv. & DS 
 &  9.88 & 14.59 & 13.56 & 15.05 & 13.27 & 19.91 & 15.45 & 13.32 & 13.49 & 15.54 \\ 
 \midrule
 DNN-PSF & PSF / LMFB & Inv. & MV 
 & 6.43 & 8.70 &  8.52 & 8.65 & 8.07 & 14.51 & 11.02 & 10.59 & 9.41 & 11.38 \\
 DNN-IBM & IBM / LMFB & Inv. & MV 
 & 6.41 & 8.63 &  8.50 & 8.39 & 7.98 & 14.28 & 11.23 & 10.39 & 9.49 & 11.35 \\
 DNN-IBM-L & IBM / LMFB + ILD & Inv. & MV 
 & 6.24 & 8.12 &  8.46 & 7.75 & 7.64 & 15.52 & 11.43 & 12.71 & 10.40 & 12.51 \\
 DNN-IBM-P & IBM / LMFB + IPD & Inv. & MV 
 & 6.52 & 8.11 & 11.41 & 8.87 & 8.65 & 14.11 & 10.81 & 11.49  & 9.47 & 11.47 \\ 
 DNN-IBM-LP & IBM / LMFB + ILD + IPD & Inv. & MV 
 & 6.54 & 8.18 &  9.53 & 8.27 & 8.13 & 15.82 & 10.63 & 12.43 & 10.25 & 12.28 \\ 
 \midrule
 MNMF-TV-WF
 & ILRMA + MNMF & Var.   & WF
 & 7.58 & 10.59 & 14.23 & 13.67 & 11.52 & 14.73 & 11.30 & 11.21 & 10.07 & 11.83 \\
 MNMF-TI-WF
 & ILRMA + MNMF & Inv. & WF  
 & 7.43 & 10.52 & 14.21 & 13.56 & 11.43 & 14.90 & 11.77 & 11.58 & 10.05 & 12.07 \\ 
 MNMF-TV-WF1 
 & ILRMA + MNMF & Var.   & WF1
 & 7.60 & 11.09 & 14.48 & 13.80 & 11.74 & 13.68 & 11.51 & 11.77 & 10.35 & 11.83 \\ 
 MNMF-TI-WF1 
 & ILRMA + MNMF & Inv. & WF1  
 & 7.68 & 11.34 & 14.53 & 13.80 & 11.84 & 14.26 & 11.54 & 11.51 & 10.24 & 11.89 \\ 
 MNMF-TV-MV
 & ILRMA + MNMF & Var.   & MV 
 & 7.71 & 11.34 & 14.61 & 14.01 & 11.92 & 14.60 & 11.73 & 11.49 & 10.12 & 11.99 \\  
 MNMF-TI-MV
 & ILRMA + MNMF & Inv. & MV  
 & 7.75 & 11.30 & 14.49 & 13.77 & 11.83 & 14.60 & 11.65 & 11.55 & 10.14 & 11.99 \\   
 \bottomrule
 \end{tabular}
\end{table*}

\begin{table*}[t]
 \centering
 \caption{The experimental results of speech enhancement (SDRs, PESQs, and STOIs) 
 for the simulated evaluation data of CHiME-3.}
 \label{tab:simu_sdr_results}
 \vspace{-2mm}
 \begin{tabular}{l|ccccc|ccccc|ccccc}
 \toprule

 & \multicolumn{5}{c|}{Simulated data (SDR)}
 & \multicolumn{5}{c|}{Simulated data (PESQ)}
 & \multicolumn{5}{c }{Simulated data (STOI)} \\
 Method
 & BUS & CAF & PED & STR & Av. 
 & BUS & CAF & PED & STR & Av.
 & BUS & CAF & PED & STR & Av.\\ 
 \midrule 
 Not enhanced
 &  6.75 &  7.74 &  8.33 &  6.56 &  7.35 
 &  2.32 &  2.09 &  2.13 &  2.19 &  2.18 
 &  0.88 &  0.85 &  0.87 &  0.86 &  0.87 \\ 
 Beamformit
 &  5.45 &  7.60 &  8.32 &  5.46 &  6.71 
 &  2.42 &  2.21 &  2.20 &  2.22 &  2.26 
 &  0.89 &  0.86 &  0.87 &  0.85 &  0.87 \\  
 \midrule
 DNN-PSF
 &  8.59 & 13.85 & 12.64 &  9.43 & 11.13
 &  2.82 &  2.52 &  2.60 &  2.61 &  2.64 
 &  0.96 &  0.94 &  0.95 &  0.94 &  0.95 \\
 DNN-IBM
 &  8.75 & 13.39 & 12.74 &  9.59 & 11.25 
 &  2.82 &  2.52 &  2.61 &  2.61 &  2.64 
 &  0.96 &  0.94 &  0.95 &  0.94 &  0.95 \\
 DNN-IBM-L
 & 10.99 & 14.38 & 12.91 & 11.12 & 12.35 
 &  2.84 &  2.54 &  2.60 &  2.63 &  2.65 
 &  0.96 &  0.95 &  0.95 &  0.95 &  0.95 \\
 DNN-IBM-P
 & 10.68 & 14.18 & 12.53 & 10.61 & 12.00 
 &  2.83 &  2.54 &  2.55 &  2.59 &  2.63 
 &  0.96 &  0.95 &  0.93 &  0.94 &  0.94 \\ 
 DNN-IBM-LP
 & 11.49 & 14.55 & 12.67 & 11.32 & 12.51 
 &  2.83 &  2.54 &  2.54 &  2.60 &  2.63 
 &  0.96 &  0.95 &  0.94 &  0.94 &  0.95 \\ 
 \midrule
 MNMF-TV-WF
 & 17.69 & 16.41 & 16.28 & 14.28 & 16.16 
 &  2.91 &  2.60 &  2.65 &  2.65 &  2.70 
 &  0.97 &  0.95 &  0.93 &  0.94 &  0.94 \\ 
 MNMF-TI-WF
 & 17.36 & 16.29 & 16.16 & 14.08 & 15.97 
 &  2.89 &  2.60 &  2.64 &  2.65 &  2.69 
 &  0.97 &  0.95 &  0.93 &  0.93 &  0.94 \\ 
 MNMF-TV-WF1 
 & 15.65 & 15.61 & 14.83 & 13.12 & 14.80 
 &  2.89 &  2.58 &  2.59 &  2.63 &  2.67 
 &  0.97 &  0.95 &  0.92 &  0.93 &  0.94 \\
 MNMF-TI-WF1 
 & 15.81 & 15.65 & 14.86 & 13.21 & 14.88 
 &  2.89 &  2.58 &  2.58 &  2.63 &  2.67 
 &  0.97 &  0.95 &  0.92 &  0.93 &  0.94 \\ 
 MNMF-TV-MV
 & 13.68 & 15.17 & 14.33 & 12.33 & 13.87 
 &  2.88 &  2.58 &  2.58 &  2.63 &  2.67 
 &  0.96 &  0.94 &  0.92 &  0.93 &  0.94 \\
 MNMF-TI-MV
 & 13.69 & 15.18 & 14.33 & 12.33 & 13.88 
 &  2.88 &  2.58 &  2.57 &  2.63 &  2.66 
 &  0.96 &  0.94 &  0.92 &  0.93 &  0.94 \\   
 \bottomrule
 \end{tabular}
\end{table*}

\subsection{Evaluation on CHiME-3 Dataset}
\label{sec:task_chime3}

We report the comparative experiment using the common dataset
 used in the third CHiME Challenge \cite{barker2015chime}.

\subsubsection{Experimental Conditions}

The training set consists of
 1,600 real utterances and 7,138 simulated ones 
 obtained by mixing the clean training set of WSJ0 with background noise.
The test set includes
 1,320 real utterances (``et05\_real\_noisy'') with $M=5$
 and 1,320 simulated ones (``et05\_simu\_noisy'') with $M=6$.
In the real data,
 each utterance was recorded by six microphones placed at a handheld tablet, 
 from which five channels except for the second channel 
 on the back side of the tablet were used
 and the fifth channel facing the speaker was set as a reference channel $m^*$. 
There were four types of noisy environments: 
 bus (BUS), cafeteria (CAF), pedestrian area (PED), and street place (STR).

To estimate TF masks,
 the five kinds of DNNs, DNN-\{IBM, PSF\} and DNN-IBM-\{L, P, LP\},
 were trained by using the simulated training set.
The DNN-HMM acoustic model was also trained using the same data.
The SDRs, PESQs, and STOIs were measured for only the simulated test set
 because the clean speech data were required.
The WERs were measured for both the simulated and real test set.

\subsubsection{Noisy Speech Recognition}
\label{sec:res_chime3asr}

The performances of ASR are listed in Table~\ref{tab:chime_wer_results}.
Among the MNMF-based variants,
 MNMF-TV-WF and MNMF-TV-WF1 
 attained the best average WERs of 11.83\% for the real data
 and were significantly better than Beamformit 
 with the average WER of 15.54\%.
Among the DNN-based variants,
 the DNN-IBM achieved 
 the best average WER of 11.35\% for the real data.
MNMF-TV-WF and MNMF-TV-WF1 were still comparable with DNN-IBM
 trained by using the matched data.
This result is considered to be promising
 because our unsupervised method does not need any prior training.
In our evaluation,
 neither the use of spatial information such as ILDs and IPDs
 nor the PSF-based cost function was effective in terms of the WER.

The WERs obtained by the MNMF-based variants
 for the simulated PED and STR data
 were worse than those for the real PED and STR data,
 while the DNN-based variants worked well for both data.
As listed in Table~\ref{tab:simu_sdr_results},
 the DNN-HMM is considered 
 to mismatch the enhanced speech for the PED and STR data
 because the performances of speech enhancement 
 for the simulated PED and STR data
 were comparable with those for the BUS and CAF data.
The WERs for the real BUS data
 were remarkably worse than those for the simulated BUS data.
A main reason would be that the spatial characteristics of speech and noise
 fluctuated over time due to the vibration of the bus in a real environment.
The low-rank assumption of MNMF still held in the bus
 and the time-variant types of beamforming thus slightly worked better.

\begin{figure}[t]
\centering
\vspace{-2mm}
\centerline{\includegraphics[width=.9\linewidth]{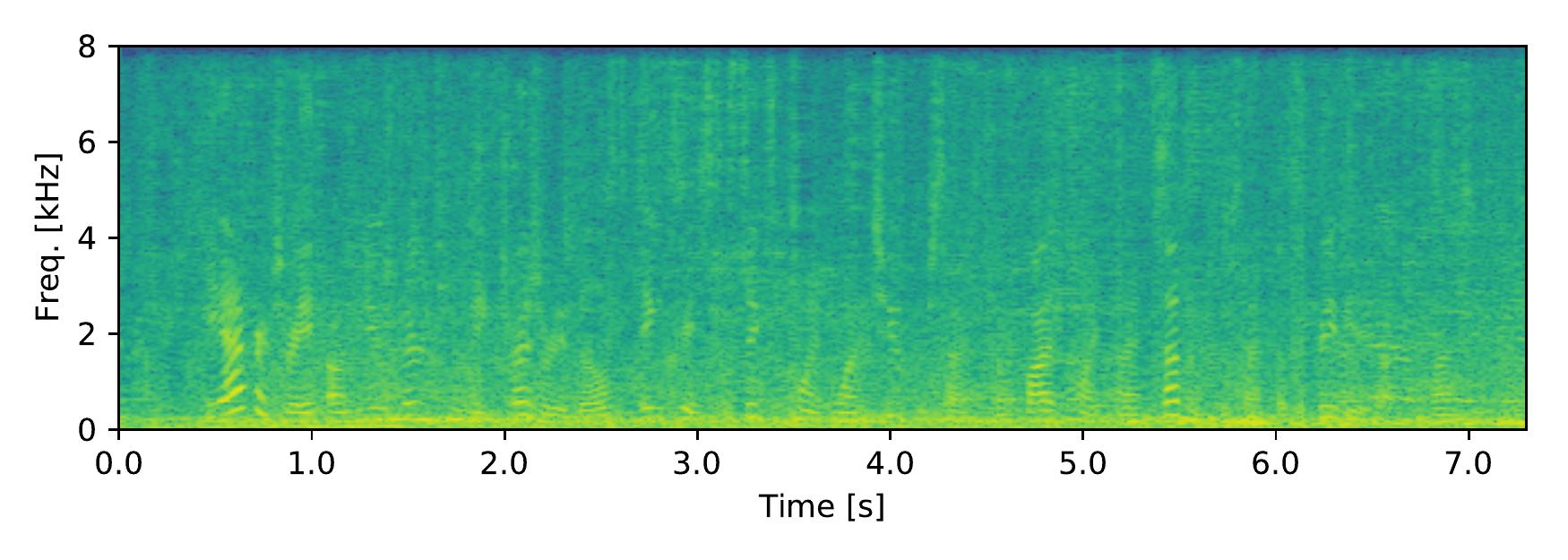}}
\vspace{-4.5mm}
\caption{The observed spectrogram of noisy speech.}
\label{fig:noisy}
\end{figure}

\begin{figure}[t]
\vspace{-2.5mm}
\centering
\centerline{\includegraphics[width=.9\linewidth]{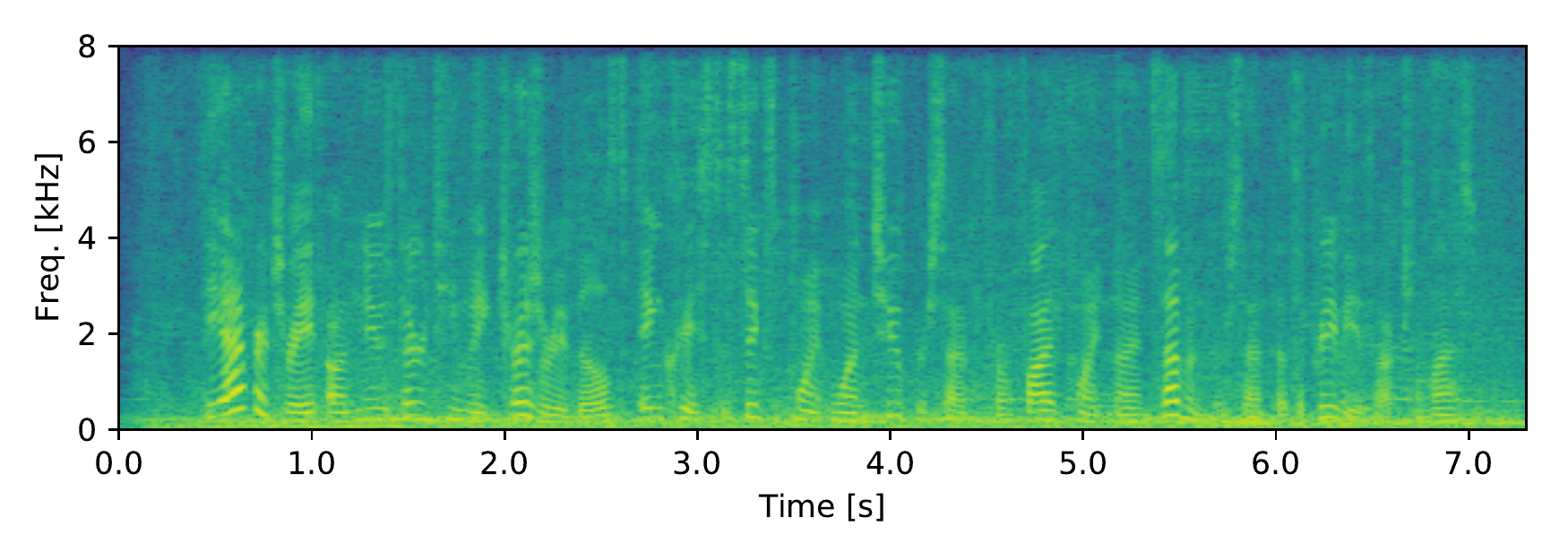}}
\vspace{-4.5mm}
\caption{The enhanced spectrogram obtained by Beamformit.}
\label{fig:beamformit_enhance}
\end{figure}

\begin{figure}[t]
\centering
\vspace{-2mm}
\centerline{\includegraphics[width=.9\linewidth]{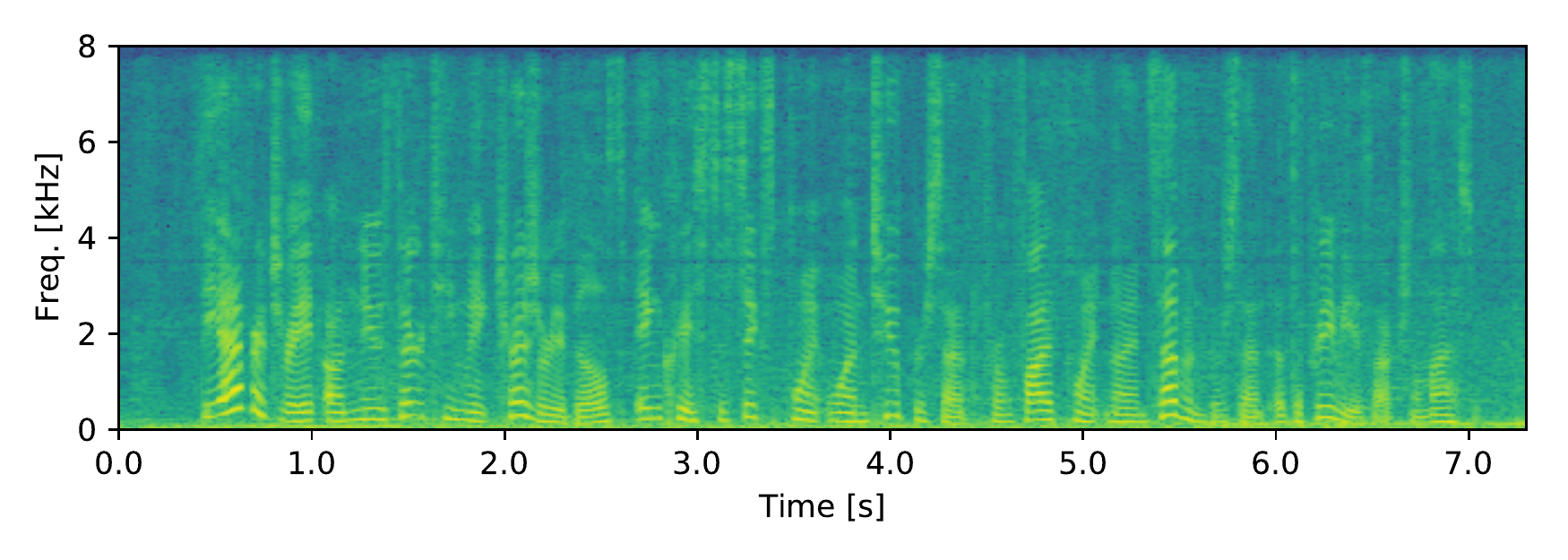}}
\vspace{-4.5mm}
\caption{The enhanced spectrogram obtained by DNN-IBM.}
\label{fig:dnnm_enhance}
\end{figure}

\begin{figure}[t]
\centering
\vspace{-2.5mm}
\centerline{\includegraphics[width=.9\linewidth]{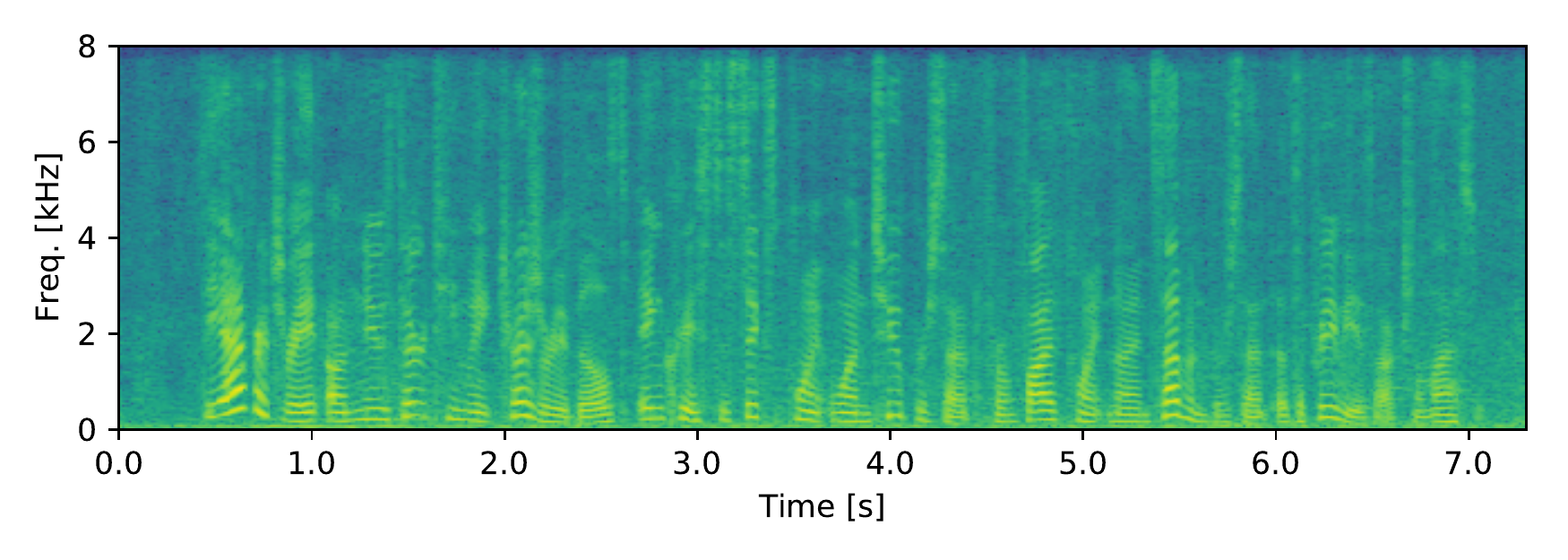}}
\vspace{-4.5mm}
\caption{The enhanced spectrogram obtained by MNMF-TI-MV.}
\label{fig:mnmf_enhance}
\end{figure}

Interestingly, 
 while the WERs obtained by the MNMF-based variants
 were much worse than those obtained by the DNN-based variants
 for the simulated data,
 all the methods yielded similar results for the real data.
This indicates that the DNN-based beamforming
 tends to overfit the training data.

\subsubsection{Speech Enhancement}

The performances of speech enhancement 
 are listed in Table~\ref{tab:simu_sdr_results}.
The MNMF-based variants 
 were generally excellent in terms of the SDR,
 and were almost comparable with the DNN-based variants in terms of the PESQ and STOI.
In our evaluation,
 the SDRs were closely related to the WERs.
MNMF-TV-WF achieved the best average SDR of 16.16 dB,
 while the DNN-based variants
 showed lower SDRs up to 12.51 dB.
Interestingly,
 the WERs obtained by the DNN-based variants
 were much better than those obtained by the MNMF-based variants
 for the simulated data.
Figures~\ref{fig:noisy}, \ref{fig:beamformit_enhance}, 
 \ref{fig:dnnm_enhance}, and \ref{fig:mnmf_enhance} 
 show the input noisy speech spectrogram
 and the enhanced speech spectrograms obtained by
 Beamformit, DNN-IBM, and MNMF-TI-MV, respectively.
Although the low-frequency noise components
 were not sufficiently suppressed by the DNN-based methods, 
 those components are considered to have a little impact on ASR.
MNMF-TI-MV was shown to estimate
 harmonic structures more clearly.

The full-rank MWF worked best in speech enhancement
 and the rank-1 MWF showed the second highest performance.
While the full-rank MWF can consider various propagation paths 
 caused by reflection and reverberation,
 the rank-1 MWF and MVDR beamforming can consider only the direct paths
 from sound sources to the microphones.
When the full-rank SCMs were accurately estimated by MNMF, 
 the performance of speech enhancement was proven to be improved.

\subsection{Evaluation on JNAS Dataset}
\label{sec:task_noisyjnas}

We report the comparative experiment using the internal dataset
 recorded in real noisy environments.
We also evaluated the online version of the proposed method.

\subsubsection{Experimental Conditions}

\begin{figure}[t]
\centering
\centerline{\includegraphics[width=0.4\linewidth]{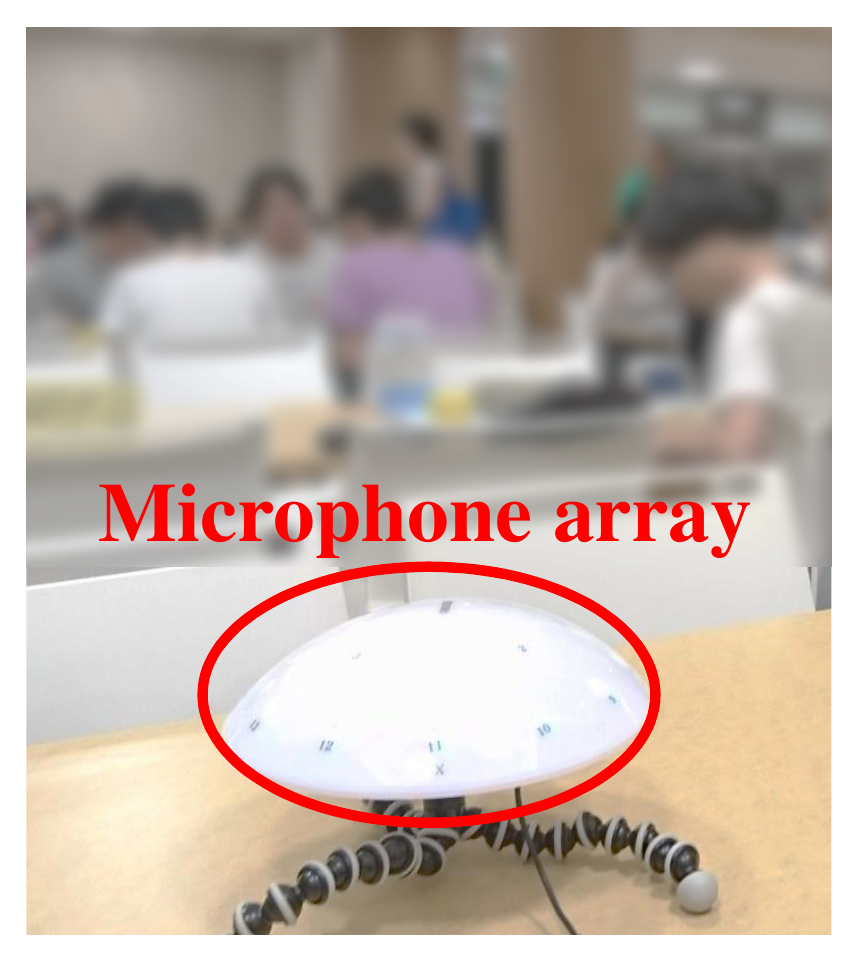}}
\caption{Recording environment for the noisy JNAS dataset.}
\label{fig:noisyJNAS}
\end{figure}

We made an internal dataset
 consisting of 200 sentences taken 
 from the Japanese newspaper article sentence (JNAS) corpus \cite{itou1999jnas} 
 and spoken by five male speakers in a noisy crowded cafeteria (Fig.~\ref{fig:noisyJNAS}).
The utterances were recorded with a five-channel microphone array ($M=5$)
 and the total duration was about 20 min.
To make a realistic condition,
 we used a hemispherical array 
 with micro-electro-mechanical system (MEMS) microphones
 that are widely used in commercial products.
The distance between the speaker and the array was 1m.
The noisy JNAS dataset has significantly different acoustic characteristics 
 from those of the CHiME-3 dataset,
 as it was recorded in a different noisy environment 
 by using a different microphone array (Table~\ref{tab:test-set}).

\begin{table}[t]
 \centering
 \caption{Comparison of two ASR tasks.}
 \vspace{-2mm}
 \label{tab:test-set}
 \begin{tabular}{lcc} \toprule
 Test set & CHiME-3 & Noisy JNAS \\ \midrule
 Noisy environments & 4~(including cafe) & 1~(another cafe) \\
 Microphone type & Condenser & MEMS \\
 Microphone array geometry & Rectangle & Hemisphere \\
 Speaker distance & 0.2 - 0.5~m & 1~m \\
 Speaker gender & 2 males \& 2 females & 5 male \\
 Speaker language & English & Japanese \\ \bottomrule
 \end{tabular}
\end{table}

The ASR performance was evaluated using this dataset.
The DNN-HMM acoustic model was also trained using the multi-condition data, 
 in which the noise data of the CHiME-3 
 were added to the original clean 57,071 utterances of the JNAS.
The model had six hidden layers with 2,048 sigmoidal nodes 
 and an output layer with 3,000 nodes.
A trigram language model was also trained using the JNAS.
The Julius decoder \cite{lee2001julius} was used in this evaluation.

The noisy JNAS task was very different from the CHiME-3 task 
 in terms of microphone setups and noise environments.
Since the DNNs for mask estimation were trained 
 using the noise data of the CHiME-3 data set,
 the noise condition of noisy JNAS was unknown.
In noisy JNAS test sets, 
 two kinds of DNNs were used for mask estimation.
One was the same DNN as that in the CHiME-3 test sets.
The other was trained by adding the noise data 
 of the CHiME-3 to the original clean utterances of the JNAS corpus.

The online versions of the proposed MNMF-based variants
 were also evaluated.
To investigate the length of the first mini-batch size,
the online enhancement processing 
 was performed using consecutive 10 utterances of the same speaker.
For online speech enhancement, the basic mini-batch size was fixed to 0.5~s,
 and experiments were conducted 
 by changing the size of the first mini-batch from 5~s to 20~s.
They were compared with the offline processing with the consecutive 10 utterances.
The value of the weight $\rho$ was set to 0.9.

\subsubsection{Noisy Speech Recognition}

\begin{table}[t]
\centering
\caption{The experimental results of noisy speech recognition (WERs) 
with offline speech enhancement for the noisy JNAS data.}
\label{tab:noisyjnas_wer_results}
 \vspace{-2mm}
 \begin{tabular}{l|c|c}
 \toprule
 Method & Training data & Avg. \\ 
 \midrule 
 Not enhanced & & 38.52 \\ 
 Weighted DS  & & 32.01 \\ 
 \midrule
 DNN-IBM & CHiME-3 & 12.27 \\
 DNN-IBM & JNAS    & 11.37 \\
 DNN-PSF & CHiME-3 & 12.11 \\
 DNN-PSF & JNAS    & 14.82 \\ \midrule
 MNMF-TV-WF  & & 10.01 \\ 
 MNMF-TI-WF  & &  9.91 \\
 MNMF-TV-WF1 & &  9.36 \\
 MNMF-TI-WF1 & &  9.30 \\
 MNMF-TV-MV  & &  9.30 \\
 MNMF-TI-MV  & &  9.36 \\  
 \bottomrule
 \end{tabular}
\end{table}

\begin{table}[t]
\centering
\caption{The experimental results of noisy speech recognition (WERs) 
with online speech enhancement for the noisy JNAS data.}
 \label{tab:wer_online_results}
 \vspace{-2mm}
 \begin{tabular}{l|c|ccccc} \toprule
 Method & Offline & \multicolumn{4}{c}{Online} \\
 First mini-batch size &  & 20 s & 15 s & 10 s & 5 s \\ 
 \midrule 
 MNMF-TV-WF  & 9.10 &  9.07 &  9.07 &  9.55 & 11.32 \\ 
 MNMF-TI-WF  & 9.17 &  9.29 &  8.98 &  9.58 & 11.56 \\ 
 MNMF-TV-WF1 & 8.94 & 10.43 & 10.72 & 10.57 & 12.21 \\ 
 MNMF-TI-WF1 & 9.00 & 10.49 & 10.95 & 10.47 & 12.30 \\  
 MNMF-TV-MV  & 8.71 &  9.95 & 11.94 & 11.21 & 12.84 \\  
 MNMF-TI-MV  & 8.78 &  9.95 & 11.68 & 10.92 & 12.78 \\   
 \bottomrule
 \end{tabular}
\end{table}

The performances of ASR are listed in Table~\ref{tab:noisyjnas_wer_results}.
Training the DNN using the JNAS data
 used for training the DNN-HMM
 was effective (from 12.27\% to 11.37\%)
 because the speech data became matched 
 in terms of spoken languages and noisy environments.
MNMF-TI-WF1 and MNMF-TV-MV achieved the best WER of 9.30\%, 
 which was 18.21\% relative improvement 
 from that obtained by DNN-IBM trained by using the data 
 used for training the DNN-HMM.
The DNN-based beamforming was found to work worse
 in unknown recording conditions.
This may have been due to over-fitting to the CHiME-3 noise data
 and it is difficult in practice to cover all the noisy conditions.

In the noisy JNAS tasks, 
 there was also only a little difference among the beamforming methods,
 but the MVDR beamforming was the most effective in combination 
 with the offline versions of the proposed method.
The use of a time-variant noise SCM also did not bring notable improvement.

\subsubsection{Online Speech Enhancement}

The performances of ASR obtained 
 by the online versions of the proposed method
 are listed in Table~\ref{tab:wer_online_results}.
The online MNMF-TI-WF achieved the average WER of 8.98\% 
 while the offline MNMF-TI-WF achieved the average WER of 9.17\%. 
The performance of the online MNMF-TI-WF using a long first mini-batch
 was better than that of the offline version
 because the initial estimates of SCMs were accurate in all frequency bins.
On the other hand, 
 the performances of the online MNMF-TI-MV and MNMF-TI-WF1
 were worse than those of the offline versions 
 even when the first mini-batch was long.
The offline MNMF-TI-MV achieved the average WER of 8.71\% 
 while the online MNMF-TV-MV achieved the average WER of 9.95\%. 
MVDR beamforming and rank-1 MWF estimated the principal eigenvector 
 of the SCM of speech as the steering vector for every mini-batch,
 which may degrade the ASR performance.
The initialization of the online versions
 depends on the first mini-batch size.
The performance was degraded when
 the first mini-batch contained a few segments of the target speech.

The practical problem of our approach
 lies in the computational complexity of MNMF
 related to the repeated inversions of SCMs.
The real-time factors of the DNN- and MNMF-based beamforming methods
 were around 0.42 and 50, respectively.
An order of magnitude faster approximations of MNMF
 \cite{ito2019fastmnmf,sekiguchi2019fastmnmf} was recently proposed, 
 which was comparable with ILRMA in speed,
 and could be extended similarly to an online version 
 for real-time noisy speech recognition.

The remaining problem lies
 in a long waiting time (10 s or 20 s) 
 before achieving reasonable performance.
This problem could be mitigated in a realistic scenario
 in which a microphone array (smart speaker)
 is assumed to be fixed in a room,
 {\it e.g.}, a microphone array is placed 
 on the center of a table for meeting recording.
Every time sound activities are detected,
 the SCMs of the corresponding directions
 can be incrementally updated.
A strong advantage of the proposed online method 
 is that is can adapt to the room acoustics on the fly.

\subsection{Experimental Findings}

The two experiments 
 using the CHiME-3 and JNAS datasets
 indicates that
 it is reasonable to use 
 the MNMF-informed time-invariant rank-1 Wiener filtering (MNMF-TI-WF1)
 for dealing with noisy speech spectrograms
 recorded in real unseen environments.
In online speech enhancement,
 the MNMF-informed time-invariant full-rank Wiener filtering (MNMF-TI-WF)
 tends to work best
 because the steering vector of speech
 is more difficult to update than the SCM of speech in an online manner.
Since the WERs and SDRs obtained by the time-invariant beamforming methods
 are almost equal to those obtained by the time-variant methods,
 in practice it would be better to use the time-invariant methods
 for improving the temporal stability of speech enhancement.

\section{Conclusion}
\label{sec:conclusion}

This paper described the unsupervised speech enhancement method 
 based MNMF-guided beamforming.
Our method uses MNMF to estimate the SCMs of speech and noise in an unsupervised manner
 and then generates an enhanced speech signal with beamforming.
We extended MNMF to an online version and initialized MNMF with ILRMA.
We evaluated various types of beamforming 
 in a wide variety of conditions.
The experimental results in real-recording ASR tasks demonstrated 
 that the proposed methods were more robust in an unknown environment 
 than the state-of-the-art beamforming method with DNN-based mask estimation.

We plan to integrate BSS- and DNN-based SCM estimation
 in order to improve the performance of ASR.
Learning a basis matrix from a clean speech database
 is expected to improve
 the performance of speech enhancement\cite{tachioka2017coupled}.
When a microphone array is specified beforehand,
 learning the normalized SCM of the target speech 
 is also expected to improve the performance.
When noisy environments are covered
 by training data used for DNN-based mask estimation, 
 MNMF can be initialized by using the results 
 of DNN-based mask estimation \cite{nakatani2017integrating}
 to further refine the SCMs of speech and noise.
It would be promising to use recently-proposed 
 semi-supervised speech enhancement methods
 based on NMF or MNMF with a DNN-based prior on speech spectra 
 \cite{bando2018icassp,leglaive2018mlsp,sekiguchi2018bayesian}.







\ifCLASSOPTIONcaptionsoff
  \newpage
\fi


\bibliographystyle{IEEEtran}
\bibliography{ieee-2018-shimada}


\begin{IEEEbiography}[{\includegraphics[width=1in,height=1.25in,clip,keepaspectratio]{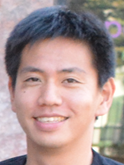}}]{Kazuki Shimada}
 received the M.S. degree in informatics from Kyoto University, Kyoto, Japan, in 2018.
He is currently working at Sony Corporation, Tokyo, Japan.
His research interests include acoustic signal processing, 
 automatic speech recognition, and statistical machine learning.
\end{IEEEbiography}

\begin{IEEEbiography}[{\includegraphics[width=1in,height=1.25in,clip,keepaspectratio]{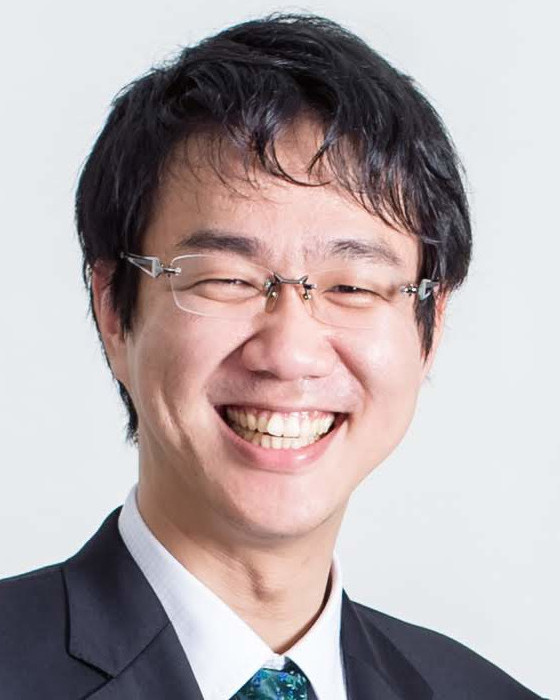}}]{Yoshiaki Bando}
 received the M.S. and Ph.D. degrees in informatics from Kyoto University, Kyoto, Japan, in 2015 and 2018, respectively.
He is currently a Researcher at Artificial Intelligence Research Center (AIRC), 
 National Institute of Advanced Industrial Science and Technology (AIST), Tokyo, Japan.
His research interests include microphone array signal processing, deep Bayesian learning, and robot audition.
 He is a member of IEEE, RSJ, and IPSJ.
\end{IEEEbiography}

\begin{IEEEbiography}[{\includegraphics[width=1in,height=1.25in,clip,keepaspectratio]{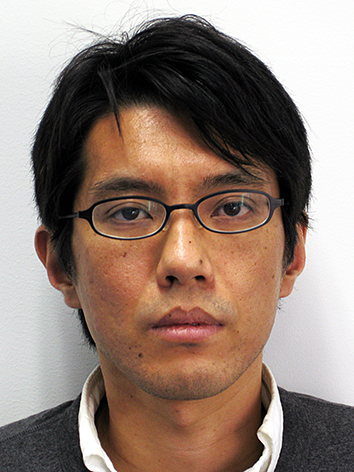}}]{Masato Mimura}
 received the B.E. and M.E. degrees from Kyoto University, Kyoto, Japan, in 1996 and 2000, respectively. 
He is currently a researcher in School of Informatics, Kyoto University. 
\end{IEEEbiography}

\begin{IEEEbiography}[{\includegraphics[width=1in,height=1.25in,clip,keepaspectratio]{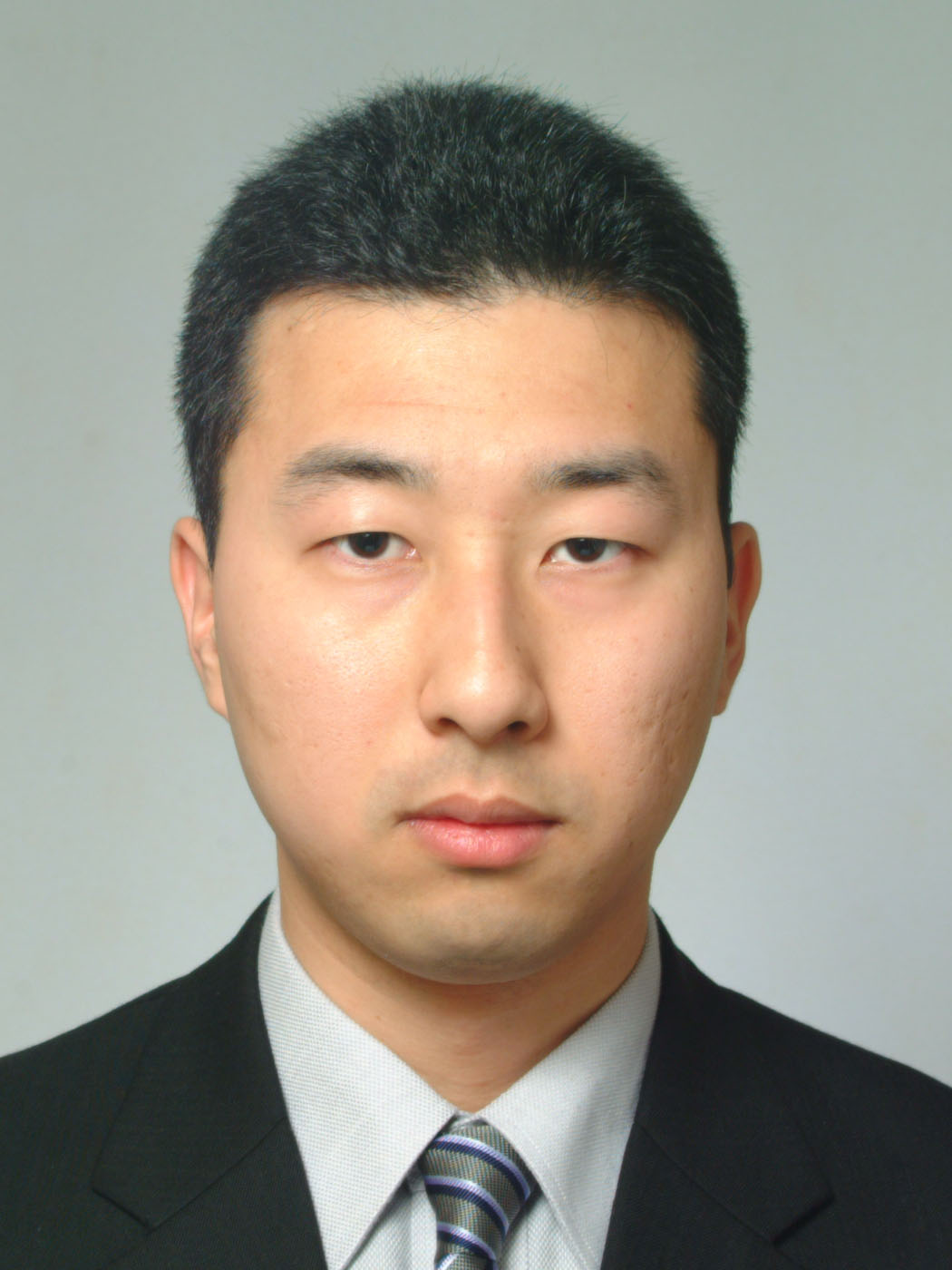}}]{Katsutoshi Itoyama}
 received the M.S. and Ph.D. degrees in informatics from Kyoto University, Kyoto, Japan, in 2008 and 2011, respectively.
He had been an Assistant Professor 
 at the Graduate School of Informatics, Kyoto University, until 2018
 and is currently a Senior Lecturer at Tokyo Institute of Technology. 
His research interests include sound source separation, 
 music listening interfaces, and music information retrieval.
\end{IEEEbiography}

\begin{IEEEbiography}[{\includegraphics[width=1in,height=1.25in,clip,keepaspectratio]{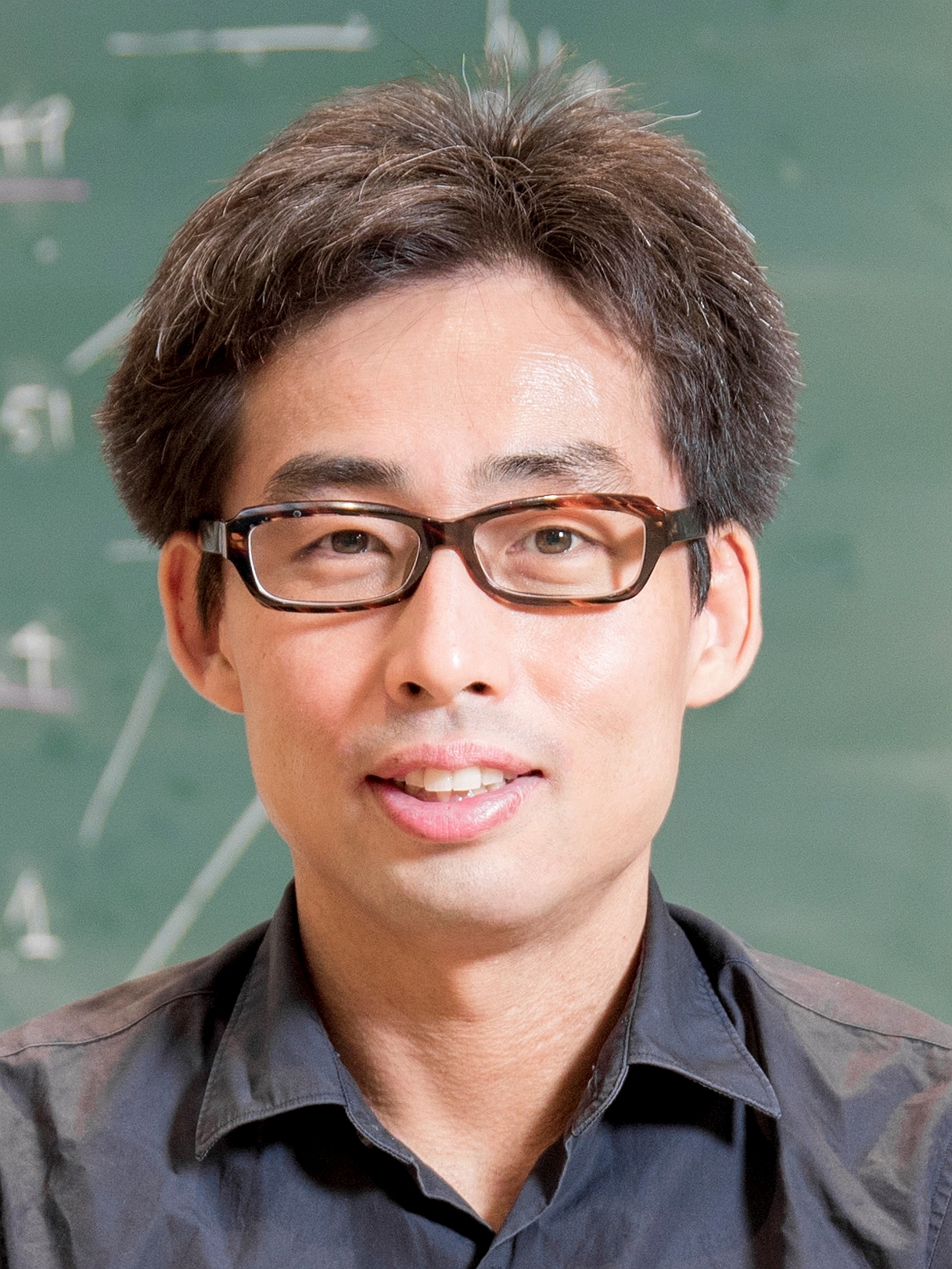}}]{Kazuyoshi Yoshii}
 received the M.S. and Ph.D. degrees in informatics from Kyoto University, Kyoto, Japan, in 2005 and 2008, respectively. 
He is an Associate Professor at the Graduate School of Informatics, Kyoto University, 
 and concurrently the Leader of the Sound Scene Understanding Team, 
 Center for Advanced Intelligence Project (AIP), RIKEN, Tokyo, Japan. 
His research interests include music informatics, 
 audio signal processing, and statistical machine learning.
\end{IEEEbiography}

\begin{IEEEbiography}[{\includegraphics[width=1in,height=1.25in,clip,keepaspectratio]{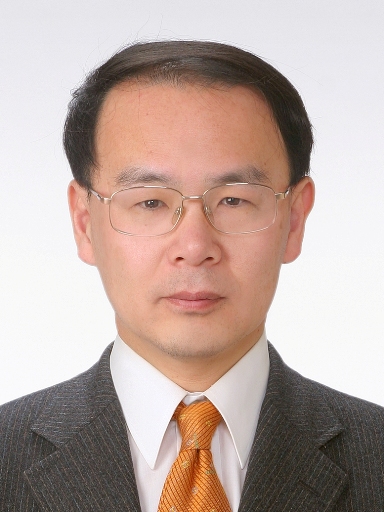}}]{Tatsuya Kawahara}
 received B.E. in 1987, M.E. in 1989, and Ph.D. in 1995, all in information science, from Kyoto University, Kyoto, Japan. 
From 1995 to 1996, he was a Visiting Researcher at Bell Laboratories, Murray Hill, NJ, USA. 
Currently, he is a Professor in the School of Informatics, Kyoto University. 
He has also been an Invited Researcher at ATR and NICT.
He has published more than 300 technical papers on speech recognition, 
 spoken language processing, and spoken dialogue systems. 
He has been conducting several projects including speech recognition software Julius 
 and the automatic transcription system for the Japanese Parliament (Diet).

Dr. Kawahara received the Commendation for Science and Technology 
 by the Minister of Education, Culture, Sports, Science and Technology (MEXT) in 2012. 
From 2003 to 2006, he was a member of IEEE SPS Speech Technical Committee. 
He was a General Chair of IEEE Automatic Speech Recognition and Understanding workshop (ASRU 2007). 
He also served as a Tutorial Chair of INTERSPEECH 2010 and a Local Arrangement Chair of ICASSP 2012. 
He has been an editorial board member of Elsevier Journal of Computer Speech and Language 
 and IEEE/ACM Transactions on Audio, Speech, and Language Processing. 
He is an editor in chief of APSIPA Transactions on Signal and Information Processing. 
Dr. Kawahara is a board member of APSIPA and ISCA, and a Fellow of IEEE.
\end{IEEEbiography}




\end{document}